\documentclass[11pt, oneside]{article} 
\pdfoutput=1

\usepackage{jheppub}
\usepackage{algorithm}
\usepackage{algpseudocode}
\usepackage{shellesc}
\usepackage{tikz}

\usepackage{tikzit}

\pgfdeclarelayer{edgelayer}
\pgfdeclarelayer{nodelayer}
\pgfsetlayers{background,edgelayer,nodelayer,main}

\tikzstyle{none}=[inner sep=0mm]
\tikzstyle{zero}=[fill=white, draw=black, shape=circle, inner sep=4pt, minimum size=0.8cm]
\tikzstyle{zeropole}=[fill={rgb,255: red,213; green,0; blue,3}, draw=black, shape=circle, inner sep=0pt, minimum size=4pt]

\tikzstyle{blue-dashed}=[-, fill=none, draw=blue, dashed]

\usepackage{amsmath,amssymb,amsfonts,amsthm}
\usepackage{svg}
\usepackage{color}
\usepackage{booktabs}
\definecolor{darkblue}{rgb}{0.1,0.1,.7}
\usepackage[]{amsmath}
\usepackage[]{graphicx}
\usepackage[]{latexsym}
\usepackage{amscd}
\usepackage[all,cmtip]{xy}
\usepackage{mathrsfs}
\usepackage{bbold}
\usepackage{subfigure}
\usepackage[margin=10pt,font=small,labelfont=bf]{caption}
\usepackage{simplewick}
\usepackage{changepage}
\usepackage{booktabs,multirow}
\usepackage[OT2,OT1]{fontenc}

\usepackage[noabbrev]{cleveref}

\newtheorem*{definition*}{Definition}
\theoremstyle{remark}

\newcommand{\data}{\text{data}}
\usepackage{nicematrix}
\NiceMatrixOptions{cell-space-limits = 4pt}
\makeatletter
\def\@fpheader{\ }
\makeatother
\usepackage{letltxmacro}
\makeatletter
\let\oldr@@t\r@@t
\def\r@@t#1#2{%
\setbox0=\hbox{$\oldr@@t#1{#2\,}$}\dimen0=\ht0
\advance\dimen0-0.2\ht0
\setbox2=\hbox{\vrule height\ht0 depth -\dimen0}%
{\box0\lower0.4pt\box2}}
\LetLtxMacro{\oldsqrt}{\sqrt}
\renewcommand*{\sqrt}[2][\ ]{\oldsqrt[#1]{#2}}
\makeatother

\title{Upgrading Extremal Flows in the Space of Derivatives}
\author{Rajeev S. Erramilli}
\affiliation{Institut des Hautes \'Etudes Scientifiques, 91440 Bures-sur-Yvette, France}
\emailAdd{erramilli@ihes.fr}

\abstract{The method of extremal flows has presented an alluring alternative approach to numerically solving bootstrap constraints. Here I present the development and adaptation of that approach to a more general class of flows with apparent discontinuities. I focus on upgrading solutions of gap maximization for the spinning modular bootstrap from low to high numerical order, though the methodology is generic to a broader class of bootstrap constraints and flows. This methodology presents various nontrivialities and nuances which reflect a richness of the space of bootstrap solutions. The result is a prototype which successfully upgrades solutions in a simple test case at small scale.}

\begin{document}

\maketitle

\section{Introduction}
\label{sec:intro}

The modern bootstrap program seeks to leverage general and non-perturbative constraints to understand the observables of quantum field theories using analytic and computational methods. Particularly the conformal bootstrap across many guises has: successfully revealed universal analytic properties of more general classes of QFT and CFT observables in state-of-the-art collider physics~\cite{Kravchuk:2018htv,Moult:2025nhu}; made contact with and helped progress problems in sphere-packing~\cite{Hartman:2019pcd,Afkhami-Jeddi:2020hde}; and provided sharp and precise numerical determinations of CFTs that were previously unimaginable. See~\cite{Poland:2018epd} for a full review of the conformal bootstrap. Focusing on numerical approaches, the bootstrap has resolved an 8\(\sigma\) tension in the literature of the 3D \(O(2)\)/XY universality class~\cite{Chester:2019ifh}, shown that the 3D \(O(3)\)/Heisenberg universality class is unstable to its cubic anisotropy~\cite{Chester:2020iyt}, and provided world-record determinations of CFT data, most notably that of the 3D Ising CFT~\cite{Chang:2024whx}. See~\cite{Rychkov:2023wsd} for a recent review on numerical developments. Most of these numerical studies have relied on interpreting the crossing equations of the bootstrap as convex optimization problems, more precisely polynomial matrix problems (PMPs) which can be expressed as semidefinite programs (SDPs), solved principally with the custom \texttt{sdpb} optimizer~\cite{Simmons-Duffin:2015qma,Landry:2019qug}. With a decade of development, \texttt{sdpb} is mature, optimized, and performant.

The most basic use of \texttt{sdpb} tries to find a dual linear functional subject to physically-motivated positivity conditions. The existence of a feasible solution proves that the bootstrap constraints cannot be satisfied, but the absence of such a functional provides little information as to what \emph{does} satisfy the bootstrap constraints. We can go further by asking \texttt{sdpb} to find not merely a feasible solution but an \emph{optimal} one with respect to an objective of our choosing. Such a solution is unique as our constraints are convex, and by strong duality there is a corresponding solution to the primal problem which gives us a truncated solution to crossing~\cite{El-Showk:2012vjm}. This primal-dual approach is robust, well-established, and has been used to great effect throughout the bootstrap literature (see the previously cited reviews for numerous examples).

Still, it leaves some things to be desired. One obvious stumbling block is the assumption of positivity, derived from the unitarity of the CFT, which obstructs the study of non-unitary CFTs.\footnote{It should be noted that there is room for some nuance. Numerical bootstrap methods have successfully (if non-rigorously) studied certain non-unitary CFTs, namely \(O(N)\) vector models with non-integer \(N>1\) and non-integer spacetime dimension \(2<d<4\)~\cite{Chester:2014gqa,Shimada:2015gda,Sirois:2022vth,Reehorst:2024vyq}. While these theories are necessarily non-unitary~\cite{Binder:2019zqc}, the non-unitarities seem to only appear in the OPE coefficients of heavy operators, and not enough to make finite-order numerics break down~\cite{Hogervorst:2014rta,Hogervorst:2015akt}. On the other hand, theories such as \(O(N)\) vector models with \(N<1\) and the Lee-Yang CFT seem to be sufficiently non-unitary so as to make standard numerical work impossible outside of the simplest single-correlator systems of constraints~\cite{Shimada:2015gda,Sirois:2022vth}.} One way around this problem is to use a strictly primal approach~\cite{Gliozzi:2013ysa,Gliozzi:2014jsa,Li:2017ukc,Hu:2025yrs,Benjamin:2026lbj}; these methods directly try to find specific truncated solutions to crossing with no dual functional, thereby bypassing the need for positivity but losing the aspect of proof.

Another stumbling block is simply performance. Owing to the necessity of arbitrary precision arithmetic\footnote{Our matrices are generically poorly conditioned for numerous reasons, requiring us to use high precision. This precision forestalls leveraging the wealth of highly-optimized implementations of linear algebra routines like BLAS and LINPACK. These libraries are the bread and butter of high performance scientific computing; through recent advances and some number-theoretic magic we can use these HPC libraries for matrix ring operations~\cite{Chang:2024whx}, but as of yet we still cannot use them for what is often the most expensive step, inverting matrices.}, \texttt{sdpb} is still very costly to run. One simple optimization is to hotstart sequential computations with prior solutions --- that is, instead of starting from scratch, one starts from a known solution that is in some sense ``nearby'' to save computing effort. This works remarkably well for feasibility computations, but for optimization computations hotstarting can lead to stalls~\cite{Rychkov:2023wsd}; this is addressed in the case of the navigator function with sky-diving~\cite{Liu:2023elz}, but it remains a stumbling block for more general optimization calculations such as sampling bounds from a grid (e.g.~\cite{Erramilli:2026ykj}). Additionally, when we increase our numerical order to refine our results, we must start from scratch no matter what, feasibility and optimization alike. This is despite the fact that carefully chosen bases in the space of functionals are expected to converge with numerical order~\cite{Mazac:2016qev}; in plain language, we're leaving information on the table.

There remains one further method not yet mentioned, namely that of so-called \emph{extremal flows}, first introduced in \cite{El-Showk:2016mxr}, which is what we will focus on in this work. In this primal-dual approach, we exchange the convex optimization problem that \texttt{sdpb} solves for a nonlinear system of equations. Given a primal-dual solution for one problem, we can solve for that of a nearby problem using conventional root-finding methods. Doing so iteratively, we can ``flow'' from one solution to another, tracking how the spectrum evolves. Positivity is not \emph{strictly} necessary to find solutions, so this method has been used to investigate non-unitary CFTs such as the Lee-Yang CFT \cite{Afkhami-Jeddi:2021iuw}; I should register that it's not immediately obvious how the solutions of this extremal approach correspond to physical non-unitary CFTs due to the loss of positivity. In any case, with extremal flows we are solving for primal-dual optimal solutions using the information of nearby solutions, so this method represents an excellent opportunity to achieve the effect of hotstarting in the case of optimal or extremal solutions were it to be made more mature and robust.

One problem of extremal flows that remains underdeveloped in the literature is that of so-called upgrading, i.e. using solutions of lower numerical order to help find solutions of higher numerical order. In simple cases such as 1D CFTs and spinless modular bootstrap, it's possible to guess the form of the higher numerical order solution and jump to an answer~\cite{El-Showk:2016mxr,Afkhami-Jeddi:2019zci}. However, for more general bootstrap problems, this becomes quite difficult, and more careful work is necessary. In this work, I will present a robust procedure for extremal flows that allows for upgrading solutions across numerical order. Along the way, I will review the method of extremal flows in a practical language before applying this procedure with a prototype code to a specific bootstrap problem, namely upgrading solutions of the spinning modular bootstrap. In doing so, I will hopefully also reveal some interesting properties of the spinning modular bootstrap as a bootstrap problem.

As there are many layers of complexity, the paper is organized so as to allow readers to skip sections if they are not interesting in that layer of technicality. \Cref{sec:review} gives a high-level overview of the extremal flow in the language we will use in the rest of the paper. \Cref{sec:theory} explains the main idea for the upgrading extremal flow. \Cref{sec:detail} provides more detail at the level of numerics and explains the procedures necessary to maintain positivity such as branch-hopping. \Cref{sec:technical} then gives details of the prototype's numerical techniques to implement those procedures. \Cref{sec:practice} applies the prototype to a specific test case, that of the spinning modular bootstrap, and presents data from a real upgrading run to make concrete what was discussed in prior sections.

\section{Flowing with extremal functionals, reviewed}
\label{sec:review}
Our ultimate goal is to find primal-dual solutions to the sorts of PMPs/SDPs that numerical bootstrappers encounter. Namely, given some space of block vectors which are functions of the conformal weights \(\vec{F}(\Delta,s)\), we would like to find a list of conformal weights \((\Delta,s)_i\) along with coefficients \(\rho_i\) and functional \(\vec\alpha\) such that
\begin{align}\label{eqn:primal}
  \vec{F}_\mathbb{1} + \sum_i \rho_i \vec{F}(\Delta_i,s_i)      & = 0,\quad \text{s.t. } \rho_i\geq 0,~\Delta_i\geq \Delta_{s_i}^\text{min};                    \\
  \label{eqn:polynomial_dual} \vec\alpha\cdot \vec{F}(\Delta,s) & \geq 0,\quad \Delta\geq \Delta^\text{min}_s,\quad \forall s\in S \subset \mathbb{Z}_{\geq 0}.
\end{align}
\Cref{eqn:primal}, the primal problem, is the classic crossing equation of the numerical bootstrap of a vector sum with positive coefficients \(\rho_i\). The inequalities \eqref{eqn:polynomial_dual} are a rephrasing of the dual optimization problem where we attempt to find a linear functional (covector) \(\vec\alpha\). In particular, if there exists a functional \(\vec\alpha\) such that the inequalities in \eqref{eqn:polynomial_dual} are strict, there cannot exist a solution to \cref{eqn:primal}. Allowing \(\vec\alpha\cdot\vec F\) to be zero, however, we can find simultaneous solutions of our equations i.e. primal-dual optimal solutions or extremal solutions. For simplicity, the case we will focus on will satisfy:
\begin{align}\label{eqn:identity_constraint}
  \vec\alpha\cdot\vec{F}_\mathbb{1}            & = 0                 \\
  \label{eqn:primal-dual}
  \rho_i\,\vec\alpha\cdot\vec{F}(\Delta_i,s_i) & = 0,\quad \forall i
\end{align}
These are a statement of one of the Karush-Kuhn-Tucker (KKT) conditions for optimal solutions to nonlinear programming. The equations above are not strictly the most general case of solutions and is specific to gap maximization problems. Other extremization problems such as OPE coefficient bounds and navigator functions will have more general expressions of \cref{eqn:identity_constraint}; I elaborate on these technical details in appendix~\ref{app:primal-dual}. It's a simple exercise to adapt the expressions throughout this work to these other extremization problems.

What do these conditions tell us? We have the requirement that each term that appears in the sum of \cref{eqn:primal} must individually be in the kernel of \(\vec\alpha\). This is trivial if \(\rho_i=0\), of course, but for the nontrivial terms where \(\rho_i>0\), we can see that
\begin{equation} \label{eqn:zero-enumeration}
  \vec{\alpha}\cdot \vec{F}(\Delta,s) = 0 \iff (\Delta,s)\in\{(\Delta,s)_i\},\quad \Delta\geq \Delta_s^\text{min}.
\end{equation}
In other words, the physically relevant zeroes of \(\alpha\cdot F\) in \eqref{eqn:polynomial_dual} are enumerated and tracked by the list of conformal weights in the sum of \cref{eqn:primal-dual} and vice versa. It was noticed in~\cite{El-Showk:2016mxr} that we can leverage this fact to describe \(\vec\alpha\) in terms of those weights/zeroes. In particular, we can trade the positivity constraint of \eqref{eqn:polynomial_dual} with equation constraints on \(\vec\alpha\cdot\vec F\) while still ensuring positivity. The logic is straightforward: a function that is positive for some \(\Delta>\Delta^\text{min}\) will be nonnegative for all \(\Delta\geq \Delta^\text{min}\) if every zero in that region has multiplicity 2, except maybe exactly at the boundary \(\Delta=\Delta^\text{min}\). So if we impose that \(\vec\alpha\cdot\vec F\) has a double zero for every \((\Delta,s)_i\) with sufficiently many double zeroes we can fix \(\vec\alpha\) completely.

\subsection{Solving the extremal flow equations, a birds-eye view}
\label{sec:birds-eye}
What does this alternative formulation buy us? We now have a nonlinear system of \cref{eqn:primal,eqn:polynomial_dual} which are solved by precisely the data \((\vec{\alpha},\{(\Delta,\rho)_i\})\). We might therefore imagine that we can sidestep semidefinite programming and instead directly solve these equations. A goal as grand as this is unfortunately quite hard. To begin, the system of equations implicitly relies upon the exact number of zeroes and their spins \(\{s_i\}\), so in fact we already need to know quite a lot about the solution before we can solve for it! That is, the form of the equations themselves is implicitly tied into the form of the solution. But even setting aside this problem, such a nonlinear system of equations will generically have a large multiplicity of solutions.

This last point might sound a bit funny, given that when solving an SDP for a primal-dual optimal solution, we are guaranteed a unique answer. The resolution to this uniqueness is that our equations have extraneous solutions. These will all violate at least one of our assumptions of positivity and will not have any correspondence with the associated SDP's solution. That is to say that in these solutions there will be \(\rho_i<0\), or \(\Delta_i <\Delta^\text{min}\), or extra zeroes of \(\alpha\cdot F\) which are not accounted for in our list. Merely solving the system of equations without imposing these positivity constraints therefore spoils the uniqueness of our solutions. Here we learn another lesson: whereas positivity is baked in and guaranteed through an SDP solver, solving this system of equations by itself does not automatically guarantee positivity. And this isn't even addressing the issue of the form of the equations i.e. knowledge of \(\{s_i\}\).

We get around the problems of multiplicity and positivity by working in a solution space with prior knowledge, near solutions that we know correspond to valid SDP solutions. This is the core idea of extremal flows~\cite{El-Showk:2016mxr}: given a solution \(\text{data}_0\) to equations \(\mathcal{E}_0(\text{data}_0)=0\) and a set of equations that are ``nearby'' \(\mathcal{E}_1(\text{data})=0\), one can find the associated solution \(\text{data}=\text{data}_1\). We expect the space of solutions to be relatively smooth in a sufficiently small neighborhood. Therefore, we can be relatively confident that the same \(\{s_i\}\) will continue to apply and that the solution \(\text{data}_1\) close to \(\text{data}_0\) will be positive; we can be confident that \(\text{data}_1\) will be the true, unique solution we're looking for. Finding such a solution can be done with any number of numerical root-finding methods, including but not limited to (quasi-)Newton methods.

As one leaves a small neighborhood of solutions into the unknown, it can happen that our solutions no longer satisfy positivity. Smoothly deformed solutions can begin to display new untracked zeroes of \(\alpha\cdot F\) which in turn would bring about regions of negativity, or more simply the coefficients \(\rho\) could become negative. We can take advantage of this violation of unitarity to seek out primal solutions to non-unitary CFTs such as the Yang-Lee minimal model~\cite{Afkhami-Jeddi:2021iuw} as mentioned before; in this case however we are deliberately abandoning positivity and the strict meaning of our dual functional. If we are after unitary CFTs, we need some prescription to maintain positivity, typically by modifying our equations to add new zeroes or to remove zeroes whose coefficients become negative~\cite{El-Showk:2016mxr,Afkhami-Jeddi:2019zci,Afkhami-Jeddi:2021iuw}. In~\cite{El-Showk:2016mxr}, the system is simple enough such that the number of zeroes of \(\alpha\cdot F\) can be predicted; in~\cite{Afkhami-Jeddi:2019zci}, the similarly simple system can have the number of zeroes guessed; and in~\cite{Afkhami-Jeddi:2021iuw}'s so-called ``unitary deformation'' zeroes are added in an ad-hoc manner to maintain positivity. In the present work, we focus on the unitary deformations, show that the procedure can be done in a systematic manner, and develop algorithms to realize the procedure. As noted in the introduction, \cref{sec:intro}, we will apply this method to a challenging test case.

A key concept to highlight here is the connection between maintaining positivity and uniqueness. A bootstrap problem necessarily satisfies positivity, and that positivity ensures that the bootstrap problem generically has a unique solution. The extremal flow equations are defined around precisely this unique, positive solution, so we are always guaranteed to have equations that can solve. However, there is no guarantee that a nearby bootstrap problem (with its own unique solution) will have the \emph{same} extremal flow equations.

However, the uniqueness, existence, and continuity of solutions means that as one system of extremal flow equations loses validity, another system with one more or fewer zero must come into validity. A way to understand it is that, on the precipice of violating positivity i.e. when one of our positivity constraints is saturated, the uniqueness of positive solutions is weakened slightly. For example, if we have a solution where \(\rho_i=0\) for some \(i\), then clearly there is a different set of equations omitting the corresponding \((\Delta,s,\rho)_i\) with materially the same solution, but with a functional with an untracked zero at what was \((\Delta,s)_i\). That is, whenever a system's solution saturates nonnegativity, there is an alternate set of equations with the same solution that also saturates a different nonnegativity. Put yet differently, if we were to start from any given primal-dual solution to a bootstrap SDP, there are cases where the exact choice of extremal flow equation is itself ambiguous; there are different branches of solutions depending on the particularities of the flows. This ``branch-hopping'' is the avatar of that condition. In short, given the uniqueness and existence of \emph{some} equation system and solution which satisfies positivity, we can only violate positivity just as another system and solution begin to satisfy positivity.

In principle, therefore, following a procedure something like the above, we can flow our way from some seed solution as far as we would like, tracking the extremal spectrum all the way.

\subsection{Practical considerations, pitfalls, and challenges}
This all paints a rosy picture, but it's worth taking a moment to take stock of the challenges that need surmounting.

Even at the beginning there's the conceptual challenge of needing a family of systems of equations which we can make sense of flowing around. There are some places where this choice is easier, however: for example, the central charge \(c\) as a free parameter for the modular bootstrap equations, as was studied by~\cite{Afkhami-Jeddi:2021iuw}, immediately gives a family of systems of equations. Without such a continuous parameter, the task becomes more challenging.

Next, there is the challenge of determining the neighborhoods we can flow in. Part of this is efficiently and effectively keeping track of positivity violations. If \(\rho<0\), that is easy to check, but checking if our functional \(\alpha\cdot F\) develops a new zero is a more difficult challenge.

Even if we can see a positivity violation, it also remains a challenge to effectively find the correct prescription to find a new branch. As we are adding and removing equations and variables and therefore changing the characteristics of our system, it's not at all guaranteed that the resulting equations can be solve in the way we expect them to.

There is one practical application which manifests all of these challenges: ``upgrading,'' i.e. flowing from a solution with \(N\) functional components (derivatives) to \(N+1\). Upgrading was investigated in~\cite{El-Showk:2016mxr,Afkhami-Jeddi:2019zci}, though they focused on simpler cases (1D correlator bootstrap, 2D spinless bootstrap) that mitigate the aforementioned challenges. The rest of this work will focus on upgrading extremal flows, to both investigate upgrading and to better understand and systematize the extremal flow procedure itself. That is, the lessons of upgrading extremal flows should readily apply to extremal flows in general.

\section{Upgrading in theory}
\label{sec:theory}
We will start with a solution to a gap maximization problem\footnote{We might instead choose a coefficient or navigator extremization problem. As discussed in Section \ref{sec:review}, there are only minor differences. For example, in this statement, we replace \((\Delta_\text{gap},\rho_\text{gap})^N\) with a coefficient \(\lambda\).}
\begin{equation} \label{eqn:gapmax}
  \text{data}^N\equiv (\vec{\alpha}^N,\{(\Delta,s,\rho)_i^N\},\{(s,\rho)_j^N\},(\Delta_\text{gap},\rho_\text{gap})^N)
\end{equation}
where \(\vec\alpha^N\in \mathbb{R}^{N}\) (though only defined up to normalization), the index \(i=1\dots K_b\) enumerates ``bulk" zeroes with freely varying scaling dimensions, and \(j=1\dots K_f\) enumerates ``fixed" zeroes which saturate their respective gap assumptions i.e. have scaling dimensions fixed to the lower bound. Explicitly, this data satisfies the following equations, using the shorthand \(F_i\equiv F(\Delta_i,s_i)\), \(\partial_\Delta F_i\equiv \partial_\Delta F(\Delta,s_i)|_{\Delta=\Delta_i}\):
\begin{align}
  \label{eqn:gap_primal}\vec F_\mathbb{1} + \rho_\text{gap} \vec{F}_\text{gap} + \sum_i \rho_i \vec{F}_i + \sum_j \rho_j \vec{F}_j & = 0                     \\
  \label{eqn:gap_bulk_dual}\vec\alpha\cdot\vec F_i = \vec\alpha\cdot\partial_\Delta\vec{F}_i                                       & = 0,\quad  i=1\dots K_b \\
  \label{eqn:gap_fixed_dual}\vec\alpha\cdot\vec F_j                                                                                & = 0,\quad  j=1\dots K_s \\
  \label{eqn:gap_gap_dual}\vec\alpha\cdot\vec F_\text{gap}                                                                         & = 0
\end{align}
These equations have the same form as equations (2.35) of \cite{Afkhami-Jeddi:2021iuw}. Note that the data also satisfies \(\vec\alpha \cdot \vec F_\mathbb{1}=0\) as an immediate consequence of all four sets of equations. To streamline ensuing discussion, let us use the following shorthand to mean the same thing:
\begin{equation}
  \label{eqn:extremal}\mathcal{E}^N(\text{data}^N) = 0.
\end{equation}
This much is just the statement of the extremal flow equations, independent of upgrading; we collect all the equations into a single expression as we will be solving all of them at once.

Allow me to briefly comment on generalities and provide some context. Solving the equations at once need not be the only strategy; in~\cite{El-Showk:2016mxr} the equations are solved in stages. In particular, bulk zeroes are subdivided into two sets, ``singles'' and ``doubles'', which differ only in at which stage all of their constraints are enforced.\footnote{The \(N\) free parameters of \(\vec\alpha\) only need as many zeroes to solve for them uniquely. However, in the present formulation, the number of zero constraints (dual equations) is equal to \(2K_b + K_f+1\), a seemingly unrelated number. We will later see in \cref{sec:jacobian_invertibility} that \(2K_b+K_f+1\geq N\). If we were to fix the other parameters and only solving for \(\vec\alpha\), this would mean we only need a subset (though not necessarily a strict subset) of the dual equations. In fact, \(\vec\alpha\) is the \((N+1)\)-covector whose kernel is spanned by \(N\) of the \(\vec F\), \(\partial_\Delta \vec F\) vectors. The remaining vectors would then necessarily be linearly dependent and unneeded for determining the functional \(\vec\alpha\). A general and reasonable choice therefore is to drop some of the \(\partial_\Delta \vec F\) vectors in determining \(\vec\alpha\). The zeroes whose \(\partial_\Delta \vec F\) vectors are used to construct \(\vec\alpha\) are referred to as ``doubles''; those whose derivative vectors are not are referred to as ``singles''. Of course, we also want to solve for the scaling dimensions of all of these vectors! So in a separate step, we must impose these remaining double zero constraints of the singles after constructing \(\vec\alpha\) also in terms of those scaling dimensions.} Roughly, we would first solve the primal equations, then construct the dual functional \(\vec\alpha\) to satisfy most of the dual equations, then solve the remaining dual zero equations; please refer to~\cite{El-Showk:2016mxr} for full technical details. This approach has benefits in interpreting the geometry of the functional \(\vec\alpha\) and the vectors \(\vec F\), \(\partial_\Delta \vec F\). In practice, it is sufficient to simply solve the primal and dual equations simultaneously at some potential numerical cost.

Returning to our upgrading problem, given a solution to \cref{eqn:extremal} and equations \(\mathcal{E}^{N+1}\) we would like to find the solution \(\text{data}^{N+1}\) such that \(\mathcal{E}^{N+1}(\data^{N+1})=0\). We would like to use our \(\text{data}^N\) as an initial guess and use root-finding to solve in the spirit of extremal flows. Already, we have a problem: \(\mathcal{E}^{N+1}\) needs one more component for \(\vec\alpha^{N+1}\), \(\alpha_{N+1}\), which we have absolutely no information about. Additionally, we have a new, independent equation which at the start will not be satisfied. In equations:
\begin{align}
  \label{eqn:primal_nplusone}F^{N+1}_\mathbb{1} + \sum_i \rho_i F_i^{N+1} & = 0 \\
  \vec{\alpha}^N\cdot \vec{F}_i + \alpha_{N+1} F_i^{N+1}                  & =0.
\end{align}
While the latter equations can be satisfied with a guess of \(\alpha_{N+1}=0\), the former equation is simply hard and will not be satisfied a priori. This matters because the principle of the extremal flow is predicated on being sufficiently close to satisfying the equations: if we start very far away, as we are here, then we are stuck.

Ultimately, this problem comes from making a discrete, discontinuous step with our equations. We therefore need some smooth deformation. Specifically, we seek a family of equations \(\mathcal{E}_\beta\)  with some auxilliary parameter \(\beta\) such that
\begin{align}
  \label{eqn:beta0}\mathcal{E}_{\beta=0}(\widetilde{\text{data}^N})                                & =0  \\
  \label{eqn:beta1}\mathcal{E}_{\beta=1}(\text{data}^{N+1}) = \mathcal{E}^{N+1}(\text{data}^{N+1}) & =0.
\end{align}
Here, \(\widetilde{\text{data}}\) signifies some embedding of the existing solution into the one-higher-derivative solution space, such as \(\alpha_{N+1}=0\) as previously mentioned. Save for the requirement that \(\widetilde\alpha\cdot F\) has the same zeroes, we purposefully leave this embedding generic as we will encounter cases where the simplest prescription will not suffice. If we had such a family of problems, we could smoothly flow across \(\beta\).

This description would seem a bit vague, admitting any number of schemes. It's therefore worthwhile to ponder what other features we might wish to see in a family \(\mathcal{E}_\beta\). As noted in prior sections, the correspondence between extremal flow equations and the corresponding convex optimization problem is key. Not only is this required for proper interpretation, the connection to the bootstrap problem guarantees us uniqueness of solution given positivity. So it would be very helpful for \(\mathcal{E}_\beta\) to correspond to a convex optimization problem for all \(\beta\), not just \(\beta=0,1\). More concretely, can we define a corresponding family \(\text{SDP}_\beta\)? This would also have the useful practical purpose of meaning that we can cross-check results with the more traditional optimization methods for all \(\beta\).

The solution is simpler than one might imagine.\footnote{I commend the reader for imagining otherwise!} Define the following:
\begin{equation}
  \label{eqn:beta_def}\boxed{\mathcal{E}_\beta^{N+1}(\text{data}) \equiv \mathcal{E}^{N+1}(\data) - (1-\beta)\mathcal{E}^{N+1}(\widetilde{\data^N}).}
\end{equation}
where we emphasize that \(\widetilde{\data^N}\) is constant and independent of the argument \(\data\). For \(\beta=1\) the RHS reduces to \(\mathcal{E}^{N+1}\), immediately satisfying \cref{eqn:beta1}. We can also see that for \(\beta=0\) \cref{eqn:beta0} is trivially satisfied.\footnote{In fact, this is trivially satisfied independent of \(\widetilde{\data^N}\); we will come to this discussion shortly.} The only explicit \(\beta\) dependence is in the coefficient \((1-\beta)\). So we now have a family of equations \(\mathcal{E}^{N+1}_\beta\) that smoothly interpolate from \(N\) to \(N+1\).

But what of the correspondence to \(\text{SDP}_\beta\)? Let's look at how the equations \eqref{eqn:beta_def} compare to the usual flow equations in more detail. Since we've assumed that the zeroes of \(\widetilde\alpha\cdot F\) are the same, the only equations that could be different for \(\mathcal{E}_\beta^{N+1}\) are \eqref{eqn:gap_primal}. They take the form
\begin{equation}
  \vec F_\mathbb{1} + \sum \rho \vec{F}  - (1-\beta)\underbrace{\left[\vec F_\mathbb{1}  + \sum \widetilde{\rho^N} \widetilde{\vec{F}^N} \right]}_{\text{constant}} = 0.
\end{equation}
where we suppress the indices and consolidate the sums for concision. The terms with superscript \(N\) are from \(\widetilde{\data^N}\) and not actually to be solved for; the entire expression in the square brackets is constant. In particular, \(\widetilde{\data^N}\) already satisfies all but the \((N+1)\)th component of \eqref{eqn:gap_primal} so the constant expression is only nonzero in the final component. We can therefore repackage all of the constant terms into a modified definition of the identity contribution which is the same as before in all but the \((N+1)\)th component:
\begin{align}
  \vec F_\mathbb{1}^\beta & \equiv \vec F_\mathbb{1} - (1-\beta)\left[\vec F_\mathbb{1}  + \sum \rho^N \vec{F}^N  \right].
\end{align}
With this modification, the equations for any \(\beta\) can be written in a form identical to the original extremal flow equations. More importantly, we can use this modified identity in the standard convex optimization approach to finding solutions; in other words, the bootstrap problems with \(\vec F_\mathbb{1}^\beta\) are precisely the \(\text{SDP}_\beta\) we seek. Therefore we also know that \emph{there is a unique, positive solution \(\data_\beta\) for every \(\beta\in[0,1]\).} In other words, this flow is completely deterministic and leaves nothing to chance or guesswork: it is exactly\footnote{There are a few unique technical subtleties which we will discuss in \cref{sec:technical}, but these subtleties are not fundamentally distinct from the subtleties that appear for any extremal flow.} like any other flow.

Allow me to comment briefly on some generalities. The \cref{eqn:beta_def} need not only apply to our upgrading problem of going from \(N\) to \(N+1\) functional components. We can in principle flow from any baseline solution to the solution of any equations \(\mathcal{E}\) this way. In fact, it might even appear that there's not a strict need that out baseline must correspond to any previous extremal solution at all. Unfortunately, things are not quite as wonderful as that. I will first note that there is a very important role that the choice of \(\widetilde{\data^N}\) plays: we expect it to satisfy our nonnegativity conditions on \(\alpha\cdot F\) so that we can smoothly flow away and protect this positivity as we flow. I will also note that having more than one equation nontrivially dependent on \(\beta\) leads to less computationally tractable problems, for reasons that will make themselves clear in \cref{sec:practice}. So in the end it seems that we really should restrict ourselves to cases where we start from a solution that is ``almost'' good by itself, like in upgrading. It's not clear to me that there would be other such cases, but it would be worth investigating nonetheless.

Returning to the problem of upgrading, we can see that we now have a family of bootstrap problems with necessarily positive and unique solutions, and therefore a corresponding family of extremal flow equations for those solutions which smoothly interpolate from \(N\) to \(N+1\) functional components. I will note that we expect that the \(N+1\)-component solution (or indeed the \(N+2\), \(N+3\)\dots) will have different (and generically more) operators appearing in their extremal spectra. This corresponds to our solutions beginning to fail positivity at intermediate values of \(\beta\) are managed by the ``branch-hopping'' alluded to in \cref{sec:birds-eye}. I leave that discussion to the following section, where we delve into the technical procedure of flowing. Along the way, we will uncover further technical subtleties and pitfalls whose resolutions will demonstrate the robustness of our procedure.

\section{Upgrading in more detail}
\label{sec:detail}
Let us return to our gap maximization problem \eqref{eqn:gapmax} now with our upgrading prescription in hand to see how the details shake out.

We begin with our variables, which is just our solution \(\data\) but as we leave the spins \(s_i\) the same, they are omitted:
\begin{equation} \label{eqn:variables}
  \data^{N+1}\to\text{vars}^{N+1}\equiv (\vec{\alpha},\{(\Delta,\rho)_i\},\{\rho_j\},(\Delta_\text{gap},\rho_\text{gap}),\beta).
\end{equation}
\(\vec\alpha\) has \(N+1\) components, but as it is only defined up to normalization, we will only count it as having \(N\) free components. We can therefore find the count of free variables:
\begin{equation}
  \# (\text{vars}^{N+1}) = N + 2K_b + K_f + 2 + 1
\end{equation}
where as a reminder \(K_b\) are the number of bulk zeroes (i.e. double zeroes with \(\Delta>\Delta_s^\text{min}\) variable) and \(K_f\) are the number of fixed zeroes (i.e. single zeroes with \(\Delta=\Delta_s^\text{min}\) fixed, not including the gap).

Next, we have the following extremal equations:
\begin{align}
  \label{eqn:upgrading_primal}\vec{F}^\beta_\mathbb{1} + \sum \rho \vec{F}(\Delta,s)          & = 0 & \text{(primal equations)}     \\
  \vec\alpha\cdot \partial_\Delta\vec{F}(\Delta_i,s_i) =\vec\alpha\cdot \vec{F}(\Delta_i,s_i) & = 0 & \text{(bulk dual equations)}  \\
  \vec\alpha\cdot \vec{F}(\Delta_j,s_j)                                                       & = 0 & \text{(fixed dual equations)} \\
  \label{eqn:upgrading_gap_dual}\vec\alpha\cdot \vec{F}(\Delta_\text{gap},s_\text{gap})       & = 0 & \text{(gap dual equation)}
\end{align}
Counting the equations:
\begin{equation}
  \# (\mathcal{E}) = N+1 + 2K_b + K_f + 1.
\end{equation}
We see that we have one more variable than equation, which means we have some one-parameter family of solutions. So if we fix one variable, for example \(\beta\), we have enough to constrain all of the other variables, no matter the number of equations and operators. Of course, we have the caveat that these are nonlinear equations, so there is actually a multiplicity of solutions. Only one will be positive and correspond to the true solution of physical interest. Hence the importance of choosing a good starting point: we start with a known positive solution for \(\beta=0\). We wish to find a healthy, positive solution for \(\beta=1\), so we find the solutions for intermediate fixed \(\beta\) as we step towards our goal. However, note that at the level of solving our system, no variable is special for us to fix: there is nothing stopping us from instead fixing a different variable and solving for \(\beta\). This perspective is useful to hold on to.

We can solve for our variables through some numerical root-finding method. Let's assume we work with some Newton step for concreteness. This involves doing iterative steps of the form
\begin{equation}
  \label{eqn:newton_step} \text{vars}\mapsto \text{vars} - \underbrace{(\mathcal{E}'(\text{vars}))}_\text{Jacobian} \!^{-1}\underbrace{\mathcal{E}(\text{vars})}_\text{error}.
\end{equation}
(Here we abuse notation and reuse \(\mathcal{E}(\text{vars})\) to mean the same as \(\mathcal{E}(\text{data})\); the meaning should be clear.) An elementary point is that this requires an invertible (non-singular) Jacobian. So while we may have the correct number of equations and variables, we must be careful to ensure that we have such a non-singular Jacobian. We will find that this puts constraints on the number of equations and operators, which we will return to in \cref{sec:jacobian_invertibility}.

But first, let us take note that in order for such a root-finding method to converge quickly in practice, we must start in some sense sufficiently close i.e. the variable values cannot change too dramatically. Practically this often means that we cannot step immediately from the \(\beta=0\) to \(\beta=1\) solutions: instead we take some sequence of intermediate \(0<\beta_1<\beta_2<\dots<1\). Along the way, we will have to ensure that our intermediate solutions correspond to the unique, positive solution of the bootstrap problem.

\subsection{Positivity constraints}
\label{sec:positivity}
As noted in \cref{sec:theory} the extremal flow equations do not themselves get us to the unique solution to the corresponding bootstrap problem: for this we need to ensure that positivity is preserved with a few extra checks. In turn, we may need to change the course of our flow to maintain positivity.

Our positivity conditions fit into two broad categories: the ``primal positivity'' inequalities on the directly physically relevant parameters \(\rho \geq 0\) and \(\Delta_i\geq s_i\), and the ``dual positivity'' of the functional \(\alpha\cdot F\geq 0\). The former are easy to check by inspection and the latter is more involved. However, since we already assume that dual positivity is saturated only for the tracked zeroes (operators) in the spectrum, it reduces to the strict inequality for all \((\Delta,s)\) outside of the tracked zeroes. I will briefly discuss the precise implementation of these checks in \cref{sec:verifying_positivity}.

If we do find that a solution fails positivity, what must we do? It's clearest to imagine an example. Suppose we are flowing from \(\beta_0\) to \(\beta_1\), and we find that for some \(i\) we have a solution with \(\rho_i <0\) at \(\beta_1\), violating primal positivity. By assumption \(\rho_i>0\) for \(\beta_0\), so by continuity we know there is some \(\beta^*\in(\beta_0,\beta_1)\) such that \(\rho_i=0\). In turn, we only have nonnegative solutions in \([\beta_0,\beta^*]\) and no further. Connecting with the intuition laid out in \cref{sec:birds-eye}, at \(\beta^*\) there is another extremal flow equation system which shares a solution. The new system differs only in omitting the constraint on the \(i\)th zero, i.e. it will have fewer equations. Since positive flows are unique, this new system necessarily will have been negative --- in fact, dual negative --- for \(\beta < \beta^*\). Without the constraint on what was the \(i\)th zero, we expect therefore that there existed some negative region in \(\alpha\cdot F\) in this neighborhood, but we will now have nonnegative solutions for \(\beta\geq \beta^*\). With this we can continue.

Extremal flows are a priori bidirectional, so we can just as easily run the story above backwards. Some system has solutions until its \(\alpha\cdot F\) begins to develop a negative region: therefore at some intermediate \(\beta^*\) there is \(\alpha\cdot F(\Delta^*,s^*)=0\). At this point there is a corresponding system in which this zero is tracked and held as a constraint, and so we can continue.

\begin{figure}[h!]
  \centering
  \includegraphics[width=0.8\textwidth]{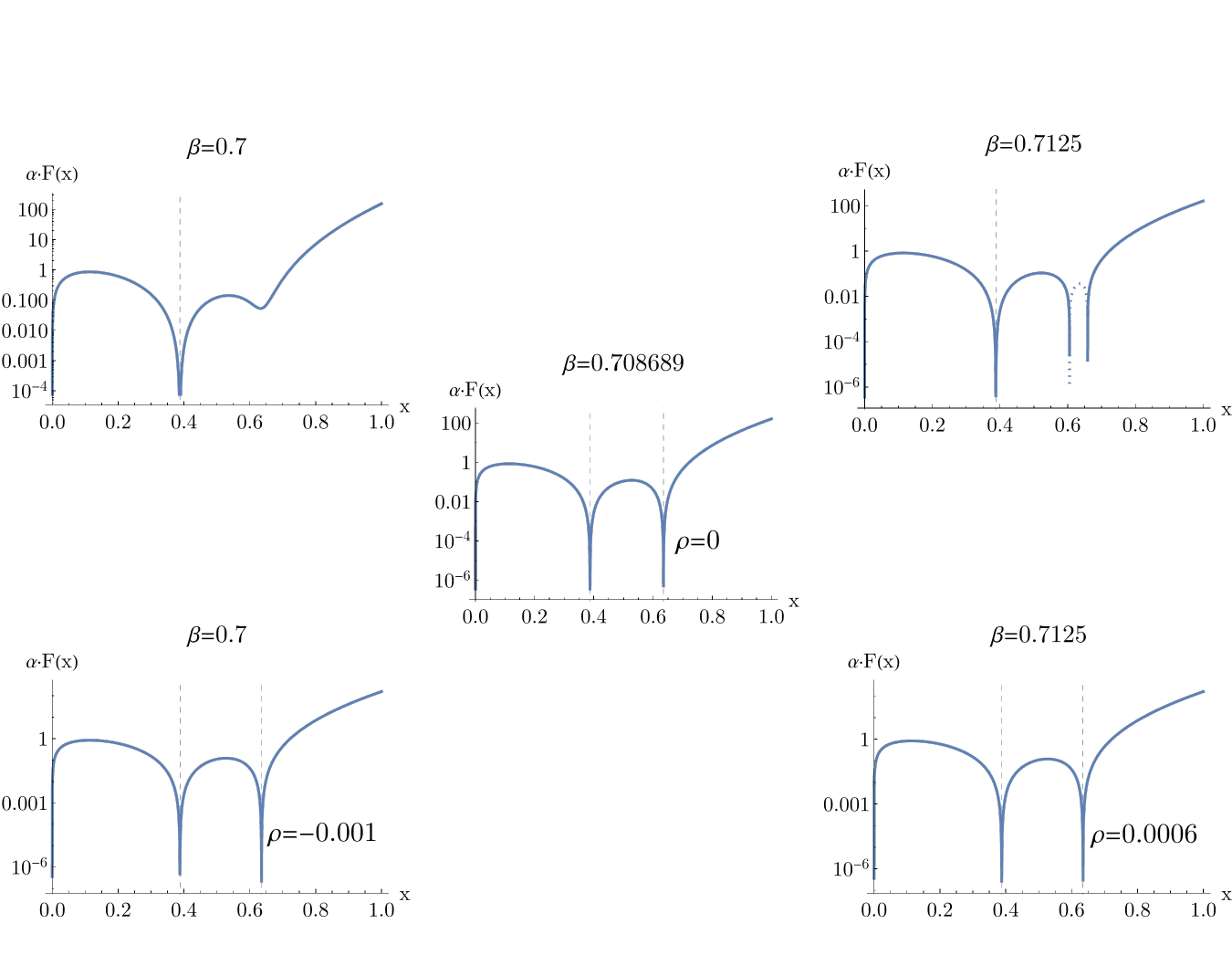}
  \caption{Plot of extremal functionals derived from real data from the flow between the \(N=19\) and \(N=20\) solutions as described in \cref{sec:results}. Each column corresponds to solutions with the same \(\beta\), but the different rows correspond to different branching paths i.e. with different tracked zeroes. Within each plot the vertical lines indicate the locations of the tracked zeroes. See main text for further comments.}
  \label{fig:branching}
\end{figure}

To make the discussion more concrete, I now present actual data showing branching solutions with positivity violations in \cref{fig:branching}. More data will be presented in \cref{sec:results}. There are two branches of solutions which cross over at the middle column. To the lower left, we see a solution with negative \(\rho\), and to the upper right we see a solution with a negative functional (indicated by the dotted lines, since we are using a log plot). The upper left and lower right are perfectly healthy and the unique ``correct'' choices for positive solutions at each value of \(\beta\). This illustrates that the upper row, where a functional develops a negative region, is complementary to the bottom row, where a coefficient \(\rho\) changes sign. Additionally, it shows the bidirectionality of this branching behavior.

The practical conclusion is that we can interpret each positivity violation as an indication to modify our equations in a prescribed way. We should be careful to note, however, that the underlying convex optimization problem remains the same. The bidirectionality of flows means these positivity violations always come in pairs. More explicitly, there are two possible versions of the discussed scenario of an operator being added/removed, for bulk and fixed zeroes respectively. If the coefficient \(\rho_i\) corresponding to a bulk zero vanishes, then we remove the constraints enforcing a double zero at \(\Delta_i\) for spin \(s_i\); if a double zero forms in our functional at \(\Delta_i\) for spin \(s_i\), then we add constraints to enforce a double zero there and include a new operator in our spectrum with \(\rho_i=0\). If the coefficient \(\rho_j\) corresponding to a fixed zero vanishes, then we remove the constraint enforcing a single zero at \(\Delta_{s_j}^\text{min}\) for spin \(s_j\); if a single zero forms at that minimum dimension, we add constraints to enforce a single zero and include a new operator in our spectrum with \(\rho_j=0\). I will summarize the relationship as follows:
\begin{align}
  \rho_i & = 0 & \rightleftarrows &  & \alpha\cdot F(\Delta_i,s_i)=\alpha\cdot \partial_\Delta F(\Delta_i,s_i) & =0  \\
  \rho_j & = 0 & \rightleftarrows &  & \alpha\cdot F(\Delta_{s_j}^\text{min},s_j)                              & =0.
\end{align}

Additionally, we have the physical constraint on bulk zeroes that \(\Delta\geq \Delta^\text{min}_s\). This is dual to the positivity constraint on fixed zeroes that \(\alpha\cdot\partial_\Delta F|_{\Delta=\Delta^\text{min}_{s_i}} \geq 0\). For a flow where a double zero reaches \(\Delta_s^\text{min}\) (and goes past), we can ``transmute'' it into a fixed single zero, leaving one zero behind at \(\Delta_s^\text{min}\) while the other goes into the unphysical region. Likewise, for a flow with a fixed zero where \(\alpha\cdot\partial_\Delta F|_{\Delta=\Delta_{s_i}^\text{min}}\) goes to zero (and goes negative), we can use the additional zero to ``transmute'' our fixed zero into a bulk double zero. Our equation summary is
\begin{align}
  \Delta_i & =\Delta_{s_i}^\text{min} & \rightleftarrows &  & \alpha\cdot \partial_\Delta F(\Delta_{s_j}^\text{min},s_j) & =0
\end{align}

Let us note that we have at this point enumerated how to violate each of the various positivity conditions in \cref{eqn:primal,eqn:polynomial_dual}, so it would appear that these are the only possible violations of positivity; I will leave open the possibility that other such problems remain. For our purposes, these are all that I have seen in practice, and each has a sensible resolution. However, there remain other obstacles along the way of flows.

\subsection{Jacobian invertibility constraints}\label{sec:jacobian_invertibility}
The Jacobian has a block structure of the form
\vspace*{2em}
\begin{equation}
  \begin{pNiceArray}[first-col,last-col]{w{c}{8em}|w{c}{8em}|w{c}{8em}}
    \sum \rho \vec{F}& 0 & \rho \partial_\Delta \vec{F} & \vec{F} & N+1 \\
    \hline
    \alpha\cdot F& \vec{F}^T & \alpha\cdot \partial_\Delta F & 0 & K_b+K_f+1\\
    \hline
    \alpha\cdot\partial_\Delta F& \partial_\Delta \vec{F}^{T} & \alpha\cdot \partial^2_\Delta F & 0 & K_b
    \CodeAfter
    \OverBrace[shorten,yshift=3pt]{1-1}{3-1}{\vec{\alpha}}
    \OverBrace[shorten,yshift=3pt]{1-2}{3-2}{\Delta_i}
    \OverBrace[shorten,yshift=3pt]{1-3}{3-3}{\rho_i}
    \UnderBrace[shorten,yshift=3pt]{1-1}{3-1}{N}
    \UnderBrace[shorten,yshift=3pt]{1-2}{3-2}{K_b+1}
    \UnderBrace[shorten,yshift=3pt]{1-3}{3-3}{K_b+K_f+1}
  \end{pNiceArray}
\end{equation}
\vspace*{1em}

The upper and left labels represent equations and variables to differentiate with; the right and lower labels count how many rows and columns there are in each block. We notice immediately that there are blocks that are identically 0. Let us first focus our attention to the last \(K_b+K_f+1\) columns, which are \((N+1)\)-component vectors padded by zeroes. If \(K_b+K_f+1>N+1\), then these would not all linearly independent, rendering the Jacobian singular. Similarly, looking at the first \(N+1\) rows, we see that if \(2K_b+K_f+2<N+1\), the Jacobian is also singular.\footnote{These same inequalities can be derived from looking at the other blocks as well.} So for the Jacobian to be non-singular, we need that
\begin{equation}\label{eqn:jacobian_n_bound}
  K_b+K_f \leq N \leq 2 K_b + K_f + 1.
\end{equation}
While this is only a necessary condition, we may expect that generically it will suffice.

It's also useful to write the inequality in an equivalent form as follows:
\begin{equation}\label{eqn:jacobian_kb_bound}
  \frac{N-K_f-1}{2}\leq K_b \leq N- K_f
\end{equation}
In this form, we have a limit on how many bulk operators there can be \(K_b\) in relation to \(N\) and the number of fixed operators \(K_f\). Given that we expect to be adding and removing operators as we flow, one might worry about the possibility that we may be obliged to remove (resp. add) an operator to maintain positivity such that we violate the lower (resp. upper) bound on \(K_b\). This would mean that our positive solution prescription becomes ill-formed, thereby leaving us seemingly stuck up a flow with no invertible Jacobian. One might hope that such pathologies don't occur in practice. Unfortunately, they do happen --- at least, those where the lower bound is violated. Even worse, it is sometimes the case that we violate this bound by attempting to lift from \(N\) to \(N+1\) components, i.e. just at the beginning of the upgrading flow. It turns out that even in these cases, it is nonetheless possible to not only continue along our flow but to do so with complete determinacy and no additional input.

What does it mean that \(K_b\) (resp. \(N\)) violates its lower (upper) bound? It means that the functional \(\vec{\alpha}\) is underdetermined -- there are too few zeroes of \(\alpha\cdot F\) to completely fix all \(N\) free components. Said a different way, there is a degeneracy of choices of \(\vec\alpha\) which all satisfy our zero equations. This degeneracy is fatal to our ability to proceed, so we need to work with this underconstrained system to find additional constraints to allow us to keep flowing. The constraints must of course be new zeroes to track.

Suppose we have exactly 1 more component of \(\vec\alpha\) than zero constraints: then we have a one-parameter family \(\vec\alpha(t)\) which satisfy our equations. In fact, for fixed \((\Delta,s)_i\) and therefore \(\vec{F}_i\) and \(\partial_\Delta \vec{F}_i\), this family is linear in our parameter \(t\):
\begin{equation}
  \vec{\alpha}(t) = \vec{\alpha}_0 + t \vec\alpha_1.
\end{equation}
The exact parametrization is arbitrary as we can always shift \(t\) as we wish. However, note that while all choices of \(t\) satisfy the zero constraints, not all will satisfy the stronger condition that the prescribed zeroes are the only zeroes i.e. that there are not other regions of nonpositivity in \(\alpha\cdot F\). It's therefore useful to choose \(\vec\alpha_0\) to be some known nonnegative functional so that \(\vec\alpha(0)\cdot F\) is nonnegative. In the case of an extremal flow where we have removed one too many operators, we have such a known \(\vec\alpha\), our existing one. In the case of the lift of \(N\) to \(N+1\), this is trivially satisfied by the existing \(\vec\alpha\) plus the choice of \(\alpha_{N+1}=0\). So we can always choose a good \(\vec\alpha_0\).

We claim that there exists a closed neighborhood of \(t\) around \(t=0\) such that \(\alpha(t)\cdot F\) is still nonnegative. In the case where we have just removed a zero from being a constraint, then \(\vec{\alpha}_0\cdot \vec F(\Delta^*,s^*)=0\) but necessarily \(\vec{\alpha}_1\cdot \vec F(\Delta^*,s^*)\neq0\). \(\vec\alpha_0\) is the unique solution once we include \((\Delta^*,s^*)\) as a zero; it was in removing this constraint that brought about the degeneracy, so \(\vec\alpha_1\) must take us elsewhere. We can assume wlog that \(\vec{\alpha}_1\cdot \vec F(\Delta^*,s^*)>0\) as we can absorb the sign into a redefinition of \(t\): this fixes us to working with \(t\geq 0\), so \(t=0\) is on the boundary of this region. The only way to develop a new zero is for there to be some \((\Delta',s')\) such that \(\vec{\alpha}_0\cdot \vec F(\Delta',s')>0\) and \(\vec{\alpha}_1\cdot \vec F(\Delta',s')<0\) so that

\begin{equation}
  \alpha(t)\cdot F<0,\qquad \text{where }t > \frac{\vec{\alpha}_0\cdot \vec F(\Delta',s')}{-\vec{\alpha}_1\cdot \vec F(\Delta',s')}.
\end{equation}

By linearity, if some region is negative for \(t=t_0>0\), then it will be negative for all \(t\geq t_0\). So therefore our functional remains positive so long as\footnote{Note the cancellation of zeroes and poles. The second expression is slightly more robust as nearly all zeroes of \(\alpha_0\cdot F\) are also zeroes of \(\alpha_1\cdot F\). Additionally, it makes sense regardless of whether \(\vec{\alpha}_1\cdot F\) has negative \emph{and} positive regions.}

\begin{equation}
  t\leq t'\equiv \min_{(\Delta,s)}\left[\frac{\vec{\alpha}_0\cdot \vec F(\Delta,s)}{-\vec{\alpha}_1\cdot \vec F(\Delta,s)}\right] = \left(\max_{(\Delta,s)}\left[\frac{-\vec{\alpha}_1\cdot \vec F(\Delta,s)}{\vec{\alpha}_0\cdot \vec F(\Delta,s)}\right]\right)^{-1}
\end{equation}.

In the case where we have a singular Jacobian when lifting from \(N\) to \(N+1\), \(t=0\) corresponds to the naive prescription of the final component of \(\vec\alpha\) being 0; this will not be generically on the boundary, so we must repeat such a procedure to find the lower bound on \(t\) as well. So we have demonstrated that we have a positive functional for \(t\in [t_-',0]\cup [0,t_+']\) for some \(t_\pm'\). By linearity, both of \(t_\pm'\) cannot be zero, so we really do have some finite neighborhood.

Let's now return to our main problem. We have an underconstrained system for \(\vec\alpha\) which is preventing us from continuing our extremal flow. We can now make this more concrete: any of \(\vec\alpha(t)\) for \(t\in [t_-',0]\cup[0,t_+']\) could work as solutions to our extremal flow equations. Pragmatically then, we need another constraint, another extremality condition to break the degeneracy: we need another zero. Zeroes come from the boundary of positivity, so I now claim that the \(\vec\alpha(t)\) we need is \(\vec\alpha_\pm\equiv \vec\alpha(t_\pm')\). In the case where \(t_-'=0\) i.e. \(\alpha_-\) is where we ``just came from'', the unique choice is then \(\alpha_+\). Adding its new zero to our list, we now have sufficiently many zeroes to completely determine \(\vec\alpha\) going forward, which indeed is part of the unique positive solution to our bootstrap problem.

In the case where \(\alpha_-\) is \emph{not} where we came from, then it would seem that even with this zero prescription we have two different possible branches to go along. However, only one of \(\alpha_\pm\) will have extremal flows in \(\beta\) increasing while maintaining positivity, as there must be a unique and positive solution. (The other will be for \(\beta\) decreasing which we are not interested in.) In particular, taking the new zeroes to be \((\Delta,s,\rho)_\pm\), only one of \(\rho_\pm>0\) will be satisfied as we continue to flow in \(\beta\), which can be checked at the level of the newly non-singular Jacobian straightforwardly. Indeed, supposing instead we wanted to flow backwards in \(\beta\) instead, then we would need to take the opposite choice in order to maintain existence.

I have verified in practice that such a pragmatic prescription seems to always work. One might be left to wonder why adding another zero is the unique resolution to this problem, even if the choice of which zero to add is unique. One way to see this is that we're performing an \emph{extremal} flow. The extrema should be unique, and so a degeneracy of extrema implies that we're not ``extremal enough''. So if we reach a point where we must remove a constraint and lose uniqueness of solution, then we need to find in that degenerate space a way to add extremality conditions and regain uniqueness. This new extremality cannot be just anything: by continuity of the flow, we need to find a condition that does not affect any of the other conditions. Remember that throughout this entire procedure that we have frozen all variables except for \(\vec\alpha\).

We might now worry about the possibility that when removing a bulk zero, we could end up underconstrained by 2. To date, we have not yet seen such a situation in a practical setting. It would be nice to understand why this doesn't seem to happen.\footnote{A speculative answer is that in such a case, the system would prefer to use its lessening of constraints to first transmute a bulk zero to a fixed zero, thereby only getting underconstrained by one.} In the case in which it does, one could also hope that we could repeat this extremality story in 2 dimensions. Conceptually it would amount to looking for kinks or cusps in the boundary of the neighborhood, but practically it would seem to be signficantly harder.

We might also worry about the case where \(K_b\) (resp. \(N\)) violates its \emph{upper} (\emph{lower}) bound. While I have not seen this in practice, it amounts to a similar story replacing the \(\vec\alpha\) components with the \(\rho\)s. With too many operators, we have degenerate ways to solve the crossing equations. Again fixing \((\Delta,s)_i\), we could repeat the above story to find \(\{\rho_i\}_\pm\) such that some \(\rho=0\) and the zero constraint can be removed. It would be nice to understand why it seems that this case is much less likely if not impossible to encounter.\footnote{One speculative answer is that the system is biased towards having fewer zeroes for whichever reason. Appealing to the geometry of functionals and vectors as elucidated in \cite{El-Showk:2016mxr}, perhaps it is simply hard to have so many vectors exist at the boundary of the kernel of the functional \(\vec\alpha\).}

In any case, it's remarkable that even in the face of the breakdown of the extremal flow procedure itself there remains a way to find the unique positive solution and keep going. The lesson seems to be that the continuity and uniqueness of positive extremal solutions should herald the direction of progress.

\section{Technical details and algorithms}
\label{sec:technical}
Having laid out the theoretical details, we now make our procedure more concrete by outlining a specific implementation. I will note that despite the choices we make in our implementation, there is no reason why this implementation should be authoritative. One should see this as a prototype.

In order to perform an upgrading extremal flow, there are four main subroutines: stepping in \(\beta\), checking the solution's positivity, ``branch-hopping'' to keep positivity, and curing Jacobian singularities. We will go into details of each, but their overall algorithm will be laid out in \cref{alg:upgrade}.

\begin{algorithm}
  \caption{The high-level algorithm of our prototype.}
  \label{alg:upgrade}
  \begin{algorithmic}
    \Procedure{upgrade}{data$_0$, $\beta_0$, $\beta_1$}
    \State eqs \(\gets \mathcal{E}_{\beta_1}\) (implicitly depends on \(\data_0\))
    \State \(\data_1 \gets \operatorname{solve}(\text{eqs},\data_0,\beta_1)\)
    \State evaluate \(\gets\) verify(data$_1$)
    \If{evaluate has length 0 (data$_1$ satisfies positivity)}
    \State \textbf{return} \(\data_1\)
    \ElsIf{evaluate has length 1 (data$_1$ violates positivity in one way)}
    \State data\(^*\), \(\beta^* \gets\) branch-hop(data\(_0\), \(\beta_0\), \(\beta_1\))
    \State \textbf{return} upgrade(data\(^*\), \(\beta^*\), \(\beta^1\))
    \Else{~(data$_1$ violates positivity in more than one way)}
    \State data\(^* \gets\) upgrade(data$_0$, $\beta_0$, $(\beta_0+\beta_1)/2$)
    \State \textbf{return} upgrade(data\(^*\), $(\beta_0+\beta_1)/2$, \(\beta_1\))
    \EndIf
    \EndProcedure
  \end{algorithmic}
\end{algorithm}

\subsection{Stepping in \(\beta\): the main flow step}
As noted, we have \(N+2K_b+K_f+3\) variables and \(N+2K_b+K_f+2\) equations, such that the space of solutions is a curve. One perspective to hold is that we have some initial solution on this curve and we would like to move along the curve to find more solutions. We can view an upgrading flow step as starting with some solution \(\data_0\) with \(\beta_0\) and solving for \(\data_1\) with \(\beta_1\). In other words, we can add an extra equation of \(\beta=\beta_1\) and solve everything with a numerical root-finding routine. For our prototype we use \texttt{FindRoot[]} in Mathematica.

Explicitly, we solve
\begin{equation}
  \begin{aligned}
    \mathcal{E}_\beta(\text{data}_\beta) & =0 \\
    \beta-\beta_1                        & =0
  \end{aligned}
\end{equation}
for the variables \(\text{vars}_\beta\).

A given set of initial and target \(\beta\) may not provide a solution which converges quickly or indeed at all. This is liable to happen when the data changes quite a lot. In practice this often happens when in between the initial and target \(\beta\) our solution branch begins to violate positivity. Therefore, if a given step fails to give a convergent solution, we simply reduce the step size. For simplicity in our prototype we recursively bisect the \(\beta\) steps.

\subsection{Verifying positivity}
\label{sec:verifying_positivity}
Verifying positivity is a critical part of our procedure for ensuring the correctness of our results. For the positivity conditions of \(\rho>0\) and \(\Delta>\Delta^\text{min}\) this is as simple as looping through the data and checking one by one.

More challenging are the requirements that
\begin{equation}
  \alpha\cdot F(\Delta,s)>0 \iff (\Delta,s)\notin \{(\Delta,s)_i\}, ~\Delta\geq \Delta^\text{min}_s.
\end{equation}
As there are continua of \((\Delta,s)\) outside of the enumerated list, it's clearly not as simple as looping and evaluating functionals.

We can solve this problem in various ways. One simple solution is to use numerical root-finding methods to find the minima of  the intervals between zeroes of the functional. That is, we loop through each such interval and search for the minima. This has the advantage of being simple, but the disadvantage of potentially getting stuck in local minima or otherwise being sensitive to the initial starting point. In other words, we have no absolute proof that a negative region isn't present between two tracked zeroes with this method.

Another method is to take advantage of the polynomial structure of \(\vec{F}(\Delta,s)\). Libraries such as MPSolve provide utilities to count, isolate, and solve for the roots of polynomials with arbitrary precision arithmetic. So we could instead use these algebraic methods to explicitly find where the zeroes of the functional are and see if there are any that are unexpected, indicating regions of negativity. These methods have the advantage of being more trustworthy in finding zeroes (we can be sure that we aren't missing regions of negativity) but have the disadvantage of complexity and potentially scaling poorly with polynomial order. In our prototype we use \texttt{CountRoots[]} and \texttt{RootIntervals[]} in Mathematica: we solve for the roots themselves using \texttt{FindMinimum[]}.

\subsection{Branch-hopping}
If we do find that our extremal flow step has failed to give us a positive solution, we must backtrack and cure this negativity by hopping onto another branch of the extremal flows. As noted before in \cref{sec:positivity}, there are places where the extremal flow equations themselves are ambiguous when coming from the unique and positive solution of an SDP: we need to make sure we pick the right branch flowing out of such places. But in order to do so, we first need to get ourselves exactly to such a point.

Suppose we are flowing from \(\beta_0\) to \(\beta_1\). We write \(\mathcal{P}(\text{data})>0\) for the positivity condition that is violated for the naive solution at \(\beta=\beta_1\). This could be \(\rho_i>0\), \(\Delta_i>\Delta_{s_i}^\text{min}\), \(\alpha\cdot F(\Delta,s)>0\), or \(\alpha\cdot \partial_\Delta F(\Delta_s^\text{min})>0\); the prescription is the same. For some \(\beta^*\in(\beta_0,\beta_1)\), we have \(\mathcal{P}(\text{data})=0\), which we can add to our system of equations. We can then solve
\begin{equation}\label{eqn:critical_beta}
  \begin{aligned}
    \mathcal{E}_\beta(\text{data}_\beta) & =0 \\
    \mathcal{P}(\text{data}_\beta)       & =0
  \end{aligned}
\end{equation}
for the variables \(\text{vars}\): we will solve for \(\beta^*\) this way.

Before proceeding, let us note that it may be the case that the equations \eqref{eqn:critical_beta} may not be solvable, either practically or more fundamentally. \(\beta_1\) being some finite interval away from the solution might lead to root-finding methods struggling, but it's possible that the violated positivity condition we identified here will not be the positivity condition \emph{first} violated coming from \(\beta_0\); in other words, we're looking for the wrong branch point. In such a case, we can bisect our step from \(\beta_0\) to \(\beta_1\) and look for positivity violations again until we find the correct one.

As an illustrative example, we found some cases where we identified a possible new single zero at a particular \(\Delta_i=\Delta^\text{min}_{s_i}\) which gave solutions that violated positivity themselves. It turned out that the correct new zero at the correct \(\beta^*\) was instead a bulk double zero at \(\Delta_i\gtrsim \Delta^\text{min}_{s_i}\).

Once we have solved for \(\beta^*\), we can proceed to the next step: altering our equations. By solving for \(\mathcal{P}(\text{data})=0\), we can be sure that we now have an extremal solution with ambiguous flow equations. Following the prescriptions from \cref{sec:positivity}, we add/remove/transmute a tracked zero in our equations and we can be sure that our solution will still satisfy this newly modified set of equations.  We can then proceed as usual in \(\beta\).

\subsubsection{Curing Jacobian singularities}

As noted in \cref{sec:jacobian_invertibility} it may be that branch-hopping or otherwise modifying our flow equations will leave us with a singular Jacobian. This needs curing before being able to proceed further. As previously noted, this is a completely deterministic procedure and there is no ambiguity or guesswork to be had. Here we will only describe the case where \(\vec\alpha\) is underconstrained (by one) as this is the only case that appeared in our worked examples, but it's straightforward to adapt this to the case where the \(\rho_i\) are underconstrained.

To summarize from before, if \(\vec\alpha\) is underconstrained, we need to search the space of solutions for a new \(\vec\alpha\) that is sufficiently constrained i.e. has a new zero.

If \(\vec{\alpha}\) is underconstrained by one, that means that the equations \(\alpha\cdot F=0\) are satisfied by an entire (affine) family of solutions
\begin{equation}
  \vec\alpha(t) = \vec\alpha_0 + t\vec\alpha_1.
\end{equation}
In order to do anything we first need to find \(\vec\alpha_0\) and \(\vec\alpha_1\). \(\vec\alpha_0\) is straightforward: as this scenario only happens when we already have an existing functional \(\vec\alpha\), we can simply take the functional that we already have. In fact, we essentially \emph{have} to: a generic solution will not satisfy all of our positivity constraints, so having a neighborhood in which we can be sure of positivity is critical. Finding \(\vec\alpha_1\) is more nontrivial: how can we be sure that this functional is not just \(\vec\alpha_0\) again? This is still straightforward. Our functionals are only unique up to an overall scaling. Practically, this means that we will often set the first component to 1. (We could in principle fix the scaling on any component!) If we then were to find a \(\vec\alpha_1\) such that its first component were 0, then the first component of \(\vec\alpha(t)\) would be 1 for all values of \(t\). Therefore we can then solve
\begin{equation}
  \begin{aligned}
    \alpha\cdot F  & =0  \\
    \alpha\cdot F' & =0  \\
    \alpha^1       & =0.
  \end{aligned}
\end{equation}
as we have as many equations as we have functional components.

Armed with \(\vec\alpha_0\) and \(\vec\alpha_1\), we can proceed as described in \cref{sec:jacobian_invertibility} and compute:
\begin{equation}
  t'_\pm = \left(\max_{(\Delta,s)}\left[\frac{\mp\alpha_1\cdot F(\Delta,s)}{\alpha_0\cdot F(\Delta,s)}\right]\right)^{-1}.
\end{equation}
Practically, we go through each value of \(s\) and find the maximum value of the function in the brackets as a function of \(\Delta\). \(\alpha_1\cdot F(\Delta,s)\) has a superset of the zeroes of \(\alpha_0\cdot F(\Delta,s)\) by construction\footnote{with the possible exception of a zero that has just been removed. However, in that case, we already know wlog that \(t'_-=0\) corresponding to this diverging maximum; for \(t'_+\)'s computation this would be a minimum and therefore a non-issue.}, so the function in the brackets is a polynomial. The function in the brackets is essentially a measure of how quickly \(\vec\alpha(t)\cdot F(\Delta,s)\) will go negative. The associated \((\Delta'_\pm,s'_\pm)\) where the function takes this maximal value is where our proposed new zero will be.

We generically keep track of two branches of solutions with \(t'_-\leq 0\leq t'_+\); by linearity \(t'_- \neq t'_+\) so they cannot both be zero. However, only one branch will have the unique solution for the ongoing extremal flow in increasing \(\beta\). In some cases we can be guaranteed that \(t'_-=0\) and that such a ``new'' zero will not give us a positive solution flowing forward. (See \cref{sec:jacobian_invertibility} for more discussion.) Otherwise, the last resort method of knowing which solution is the correct one is straightforward. Since we are adding a new tracked zero, there is a new variable \(\rho'\) associated with that zero that will be 0. We only need see whether for increasing \(\beta\) this variable immediately trends negative or positive. We can do so with the Jacobian, or equivalently with a Newton step. Schematically, we ask:
\begin{equation}
  \frac{d \rho'}{d\beta} \overset{?}{>}0.
\end{equation}
In practice I have verified that this procedure seems to always work, lending credence to our intuiton for existence and uniqueness of solution.

\section{Upgrading in practice: Spinning modular bootstrap}
\label{sec:practice}
We now pass to practical details of the specific bootstrap problem we apply our methods to, namely the spinning modular bootstrap. Note that while much of our procedure is generic, there are some details specific to the modular bootstrap.
\subsection{Modular Bootstrap Re-revisited}

Let us first quickly review the basic details of the modular bootstrap. I will elide over details for brevity and direct the reader to prior literature for full explanations \cite{Collier:2016cls,Afkhami-Jeddi:2019zci}.
The torus partition function of a 2D CFT has the form
\begin{equation}
  Z(\tau) = \sum_{h,\bar{h}} m(h,\bar{h}) \chi_h(\tau) \chi_{\bar h}(\bar{\tau})
\end{equation}
where \(m(h,\bar{h})\) are the multiplicities of the representations \((h,\bar{h})\) and \(\chi_h(\tau)\) are the Virasoro characters as a function of the modular parameter \(\tau\). For a Verma module,
\begin{equation}
  \chi_h(\tau) = \eta^{-1}(\tau) \,q^{h-\frac{c-1}{24}},
\end{equation}
where \(q=e^{2\pi i\tau}\) is the nome. We will assume no shortened multiplets except that of the identity (including the stress tensor) where \(h=\bar h=0\). The torus partition function is modular invariant, i.e. \(Z(\tau)=Z(-1/\tau)\), which nontrivially constrains both the representations that appear as well as their multiplicities, though somewhat obscurely. We can make headway with the constraints by combining terms into ``modular blocks'' to get
\begin{equation}\label{eqn:crossing}
  F_\mathbb{1}(\beta,\bar\beta)+\sum_{h,\bar h} a_{h,\bar h} F_{h,\bar h} (\beta,\bar\beta) = 0
\end{equation}
expressed in the variable \(\beta=i\tau\).\footnote{In this variable, crossing is \(\beta\leftrightarrow \beta^{-1}\).} For full details involved in this reduction, see \cite{Collier:2016cls,Afkhami-Jeddi:2019zci}. Here we have separated the identity from the other more general representations in the sum. Explicitly,
\begin{align}
  F_{h,\bar h} (\beta,\bar\beta) & = (\beta\bar\beta)^{1/4} \left[e^{-2\pi\left[\beta h+\bar\beta\bar h - (\beta+\bar\beta)\frac{c-1}{24}\right] }+ (h\leftrightarrow \bar h)\right] + (\beta\leftrightarrow \beta^{-1}) \\
  F_\mathbb{1}(\beta,\bar \beta) & = F_{0,0}(\beta,\bar\beta) - 2F_{1,0}(\beta,\bar\beta) + F_{1,1}(\beta,\bar\beta).
\end{align}
At this point, to convert the equation \eqref{eqn:crossing} into a numerical constraint, we need a suitable basis of functionals. A particularly nice one uses the following differential operators \cite{Afkhami-Jeddi:2019zci}:
\begin{align}
  \mathcal{D}^p f(\beta)                                        & =\frac{1}{p!}\left[\frac{(\beta+1)^2}{2}\frac{\partial}{\partial\beta}\right]^p\left[\sqrt{\frac{\beta+1}{2\sqrt\beta}} f(\beta)\right] \\
  \mathcal{D}^p(\beta^{1/4}e^{-\beta A}) |_{\beta=1}            & = L_p^{(-1/2)}(2A)\,e^{A}                                                                                                               \\
  \mathcal{D}^p(\beta^{1/4}e^{-\beta A})|_{\beta\to \beta^{-1}} & =  (-1)^p \mathcal{D}^p(\beta^{1/4}e^{-\beta A}).
\end{align}
where we notice that the differential operators evaluated at the modular-symmetric point \(\beta=1\) yield generalized Laguerre polynomials, and that around that point the action of crossing \(\beta\to\beta^{-1}\) is quite simple. With these differential operators in mind, we can then choose a functional basis
\begin{align}
  \alpha[F] & = \sum_{p,\bar p} \alpha_{p,\bar p} \,\mathcal{D}^p\, \bar{\mathcal{D}}^{\bar p} \,F(\beta,\bar\beta) \big |_{\beta=\bar\beta=1}
\end{align}
which takes advantage of these nice properties. Explicitly now in terms of the original blocks we can define
\begin{align}
  F^{p,\bar p}_{h,\bar h} & \equiv \mathcal{D}^p\, \bar{\mathcal{D}}^{\bar p} \,F(\beta,\bar\beta) \big |_{\beta=\bar\beta=1} = \mathcal{L}_p(h)\mathcal{L}_{\bar p}(\bar h) + (h\leftrightarrow\bar h) \\
  F^{p,\bar p}_\mathbb{1} & \equiv F^{p,\bar p}_{0,0} -2e^{-2\pi} F^{p,\bar p}_{1,0} + e^{-4\pi} F^{p,\bar p}_{1,1}                                                                                     \\
  \mathcal{L}_p(h)        & \equiv L_p^{(-1/2)}\left[ 4\pi\left(h-\frac{c-1}{24}\right)\right].
\end{align}
Note that our functions are even in \(p\leftrightarrow\bar p\).
With this suitable basis of functionals defined, we proceed to how it translates into the numerical problem we go about solving. We consider a truncation of the space of derivatives, of which we only need to consider the ones that are odd:
\begin{multline}
  \{(p,\bar p) ~|~ p+\bar p \in 2\mathbb{N}-1,~p+\bar p\leq \Lambda\} = \{(1,0),(3,0),(2,1)\dots\}\\
  \cong \{n~|~ n\in\mathbb{N},~n\leq \frac{1}{8}(\Lambda+1)(\Lambda+3)\}.
\end{multline}
Here the truncation is parametrized by \(\Lambda\) and we have reindexed to a single vector index \(n\) for convenience; we will often suppress this index.
Exchanging \((h,\bar h)\) for \((\Delta,s)\) and appropriately rescaling the coefficients, we have crossing equations of the form
\begin{equation}\label{eqn:truncated_crossing}
  \vec{F}_\mathbb{1} + \sum_{\Delta,s} \rho_{\Delta,s} \vec{F}_{\Delta, s} = 0.
\end{equation}
We have now arrived at a form equivalent to what was discussed in prior sections.\footnote{Here our \(\vec F_{\Delta, s}\) is equal to \(\mathcal{F}^i_{\Delta,s}\) in \cite{Afkhami-Jeddi:2021iuw}.}

However, this form as-is posed various technical difficulties for what we wish to do. In the following sections, we will make various tweaks to this form to make it more numerically tractable. The overall form, however, will remain the same. Where it's relevant, we will use different symbols to represent the scaled and tweaked objects; where it is not, we will stick with the symbols \(\vec{F}\). I hope it will be clear in practice.

\subsection{The topology of modular blocks}
\label{sec:block_topology}
In our initial attempts to upgrade modular bootstrap problems, we discovered a series of worrying behaviors. First, we noticed that sometimes the spectra given by
\texttt{spectrum-extraction} were less than complete, in the sense of being unable to satisfy the crossing relations.\footnote{In general numerical results only satisfy the extremal flow equations up to an epsilon that's necessary in the numerics as discussed in \cite{El-Showk:2016mxr}; it is quickly resolved by solving the extremal flow equations. The issue discussed here is an additional one.} Upon manual inspection, we came to discover that there were additional zeroes at very large \(\Delta\) that were not being appropriately detected. Separately, we came to notice the occasional presence of operators with very large spin at or near our cutoff \(s_\text{max}\). These were accompanied by our functional becoming negative immediately after the cutoff, indicating a cutoff-dependent behavior of our solutions and that our solutions were not actually positive. Practically, this is not a procedural problem if this dependence is mild, though certainly it draws some uncertainty as to the validity of our solutions.

We came to realize that both of these problems are related to an interesting fact: it is useful to consider not just the standard modular blocks but those ``at infinity'', i.e. in the limit of very large twist and/or spin. In other words, we need in some sense the closure of the space of modular blocks. The infinite-dimension limit of blocks at finite spin \(s=|h-\bar h|\) is easiest managed by compactifying the space of twists \(2\bar h = \Delta-s\in[0,\infty)\) to \(x\in[0,1]\) via the relation
\begin{equation}
  \bar h = \frac{x}{1-x}.
\end{equation}
However, Laguerre polynomials diverge in the infinite limit:
\begin{equation}
  L_{p}^{(\alpha)}(z) \overset{z\to\infty}{\sim} \frac{(-z)^p}{p!}.
\end{equation}
so we expect that in the compact \(x\) coordinate there will be a corresponding pole \((1-x)^{-p-\bar p}\) which is numerically problematic for our procedure. (i.e. How can the crossing equation have a term with a nonzero coefficent that is divergent?) Here I note that we are free to make \(x\)-dependent rescalings of the blocks \(F\) so long as we account for it in the positive coefficients \(\rho\) thereby keeping the product \(\rho F\) the same. Using the explicit expression of Laguerre polynomials and a few lines of algebra, we arrive at the new closed form expression
\begin{align}
  \widetilde{F}^{p,\bar p;\Lambda}_{x,s} & \equiv (1-x)^{\Lambda-p-\bar p}\widetilde{\mathcal{L}}_p\left(x,-\frac{c-1}{24}\right)\widetilde{\mathcal{L}}_{\bar p}\left(x,-\frac{c-1}{24}+s\right) + (p\leftrightarrow \bar p) \\
  \widetilde{\mathcal{L}}_p(x,b)         & \equiv \sum_{k=0}^p \frac{(-4\pi)^k}{k!}\binom{p-1/2}{p-k} (x+(1-x)b)^k\,(1-x)^{p-k}.
\end{align}
Note that because the blocks diverge as a function of the order of their derivative, we need to specify the derivative order truncation \(\Lambda\geq p+\bar p\) in this definition for all the components. These new blocks are related to the old blocks by the relations
\begin{equation}
  F^{p,\bar p}_{h,\bar h} = (1-x)^{-\Lambda} \widetilde{F}^{p,\bar p; \Lambda}_{x,s};\qquad \bar h =\frac{x}{1-x} = h-s.
\end{equation}
Factoring out the essential singularity of a polynomial at infinity also has the nice effect of ``evening out'' the block as a function of dimension which we observed made the various numerical parameters much closer in order of magnitude and made our calculations more stable. We can now see the form of the infinite-twist limit.
\begin{equation}
  \widetilde{F}^{p,\bar p;\Lambda}_{1,s} = \begin{cases}
    0                                     & p+\bar p < \Lambda    \\
    2\frac{(-4\pi)^{\Lambda}}{p! \bar p!} & p + \bar p = \Lambda.
  \end{cases}
\end{equation}
Interestingly, the \(s\)-dependence of the blocks has dropped out completely, and the only nonzero elements are at the highest derivative order. Additionally, the blocks are strictly negative in their nonzero components.

We will ultimately be interested in single and double zeroes of functionals over these blocks, so for our purposes we also need the derivative of the block at \(x=1\). This is
\begin{equation}
  \partial_x \left.\widetilde{F}^{p,\bar p;\Lambda}_{x,s}\right|_{x=1} = \begin{cases}
    0                                                                                                                                     & p+\bar p < \Lambda    \\
    \frac{(-4\pi)^{\Lambda-1}}{p! \bar p!}\left[ p\left(1-2p+4\pi\left[s-\frac{c-25}{12}\right]\right) + (p\leftrightarrow \bar p)\right] & p + \bar p = \Lambda.
  \end{cases}
\end{equation}
Here we can see that there is a (linear) spin dependency, and indeed one that changes sign. However, we can rearrange terms to get
\begin{equation}
  \partial_x \left.\widetilde{F}^{p,\bar p;\Lambda}_{x,s}\right|_{x=1, p+\bar p=\Lambda} = 4\frac{(-4\pi)^{\Lambda-1}}{(p-1)! (\bar p-1)!} - \frac{\Lambda}{2}\left(s-\frac{c-25}{12}+1-2\Lambda\right)\widetilde{F}^{p,\bar p;\Lambda}_{x,s}.
\end{equation}
In other words, the only spin-depedent terms are independent of \(p,\bar p\) except proportionally to \(\widetilde{F}\) itself. We only need \(\partial_x\widetilde{F}\) in conjunction with \(\widetilde{F}\) itself, so we can effectively project out the proportional term and be left with an expression that is strictly \(s\)-independent again. So across all values of \(s\), they develop single and double zeroes together: all of the families coincide at a junction at \(x=1\).

But what of the limit of infinite \emph{spin}? Or rather, the limit of infinite spin with a fixed ratio between twist and spin. Such a family is not strictly necessary per se, but the presence of operators at or near the cutoff \(s_\text{max}\) implies some behavior at large spin; considering this limit of infinite spin (holding the ratio of twist and spin fixed) would help ameliorate our cutoff dependencies in our solutions. The twist/spin ratio can be anywhere in \([0,\infty)\), much like the twist itself in the finite spin case, and with a suitable compactfication we arrive at
\begin{equation}
  \widetilde{F}^{p,\bar p;\Lambda}_{x,\infty} \equiv \delta_{p+\bar p,\Lambda} \frac{2(-4\pi)^{\Lambda}}{p! \bar p!} x^{\min(p,\bar p)}\sum_{k=0}^{|\bar p-p|/2}\binom{|\bar p-p|}{2k}(1-x)^k.
\end{equation}
Note that the \(x\) here is not precisely the same \(x\) used in the finite spin cases: however, we feel the abuse of notation is justified in that we can immediately see that the expressions coincide for \(x=1\). The first derivative doesn't precisely coincide:
\begin{equation}
  \partial_x \left.\widetilde{F}^{p,\bar p;\Lambda}_{x,\infty}\right|_{x=1} = \delta_{p+\bar p,\Lambda}\frac{(-4\pi)^{\Lambda}}{p!\bar p!} \left[\Lambda(1-\Lambda) + 4p\bar p\right] = 4\delta_{p+\bar p,\Lambda}\frac{(-4\pi)^{\Lambda}}{(p-1)!(\bar p-1)!} + \frac{\Lambda(1-\Lambda)}{2}\widetilde{F}^{p,\bar p;\Lambda}_{x,\infty}
\end{equation}
Nonetheless we see that it is a linear combination of the finite-\(s\) derivative and the \(\widetilde{F}\) itself, so it is not independent. Notably, however, when projecting out \(\widetilde{F}\), we have the opposite sign relative to the finite-\(s\) derivative. This means that a single zero at \(x=1\) would necessarily have a negative region either for finite spin or infinite spin, and thus assuming all families of blocks are present the only positive possibilities are to either have no zero or a double zero at \(x=1\).

We should take note of the fact that we will be working with finite truncations of our block vectors. There is a degenerate case of only one element of \(\widetilde{F}\) with \(p+\bar p=\Lambda\), so in turn all of these block vectors at infinity are degenerate. In turn, there is no family parametrized by \(x\) at \(s=\infty\), and zeroes at \(x=1\) \emph{can} be single zeroes.

\begin{figure}[h!]
  \centering
  \hspace{2cm}
  \begin{tikzpicture}
    \begin{pgfonlayer}{nodelayer}
      \node [style=zero] (0) at (0, 0) {0};
      \node [style=zero] (1) at (0, 1) {1};
      \node [style=zero] (2) at (0, 2) {2};
      \node [style=zero] (3) at (0, 3) {3};
      \node [style=zero] (4) at (0, 4) {4};
      \node [style=none] (5) at (2, 4) {};
      \node [style=none] (6) at (2, 3) {};
      \node [style=none] (7) at (2, 2) {};
      \node [style=none] (8) at (2, 1) {};
      \node [style=none] (9) at (2, 0) {};
      \node [style=zero] (10) at (8, 0) {};
      \node [style=zero] (11) at (0, 7) {$\infty$};
      \node [style=none] (13) at (0, 5.5) {\vdots};
      \node [style=none] (14) at (0,-0.7) {x=0};
      \node [style=none] (15) at (8,-0.7) {x=1};
    \end{pgfonlayer}
    \begin{pgfonlayer}{edgelayer}
      \draw [style=black] (0) to (9.center);
      \draw [style=black] (1) to (8.center);
      \draw [style=black] (2) to (7.center);
      \draw [style=black] (9.center) to (10);
      \draw [style=black, in=180, out=0, looseness=1.25] (8.center) to (10);
      \draw [style=black, in=180, out=0, looseness=1.25] (7.center) to (10);
      \draw [style=black, in=-180, out=0] (6.center) to (10);
      \draw [style=black, in=-180, out=0, looseness=0.75] (5.center) to (10);
      \draw [style=black] (3) to (6.center);
      \draw [style=black] (5.center) to (4);
      \draw [style=blue-dashed, in=0, out=0] (11) to (10);
    \end{pgfonlayer}
  \end{tikzpicture}
  \caption{The topology of the space of modular blocks. On the left are the \(x=0\) boundaries of each of the block families, each labeled by a spin \(s\) in the circles. The finite-\(s\) block families all meet at a junction of \(x=1\); the \(s=\infty\) family meets on the other side of the junction due to having the opposite sign of derivative. The \(s=\infty\) family is depicted as a dashed, blue line as it only exists for truncations where there are more than one component of the block vector such that \(p+\bar{p}=\Lambda\). See main text for further details.}
  \label{fig:block_topology}
\end{figure}
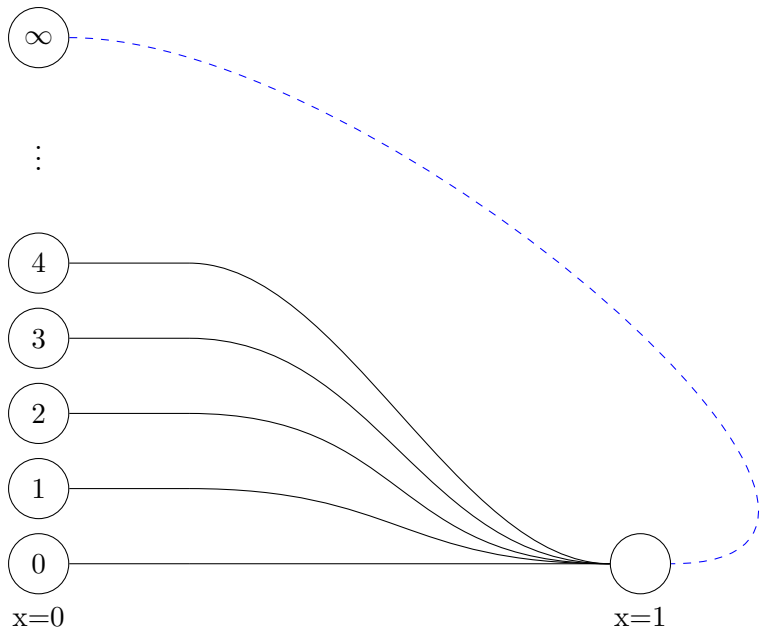

We arrive then at the full picture of blocks. Without the degeneracy, we have \(s_\text{max}+2\) independent families of blocks parametrized over the interval \([0,1]\). They all have their own simple boundaries at \(x=0\), but they share a common junction at \(x=1\) which cannot support a single zero, but can allow for the ``transit'' of double zeroes from finite to infinite \(s\) and vice versa. With the degeneracy (when there is only one component with \(p+\bar p = \Lambda\)), we have \(s_\text{max}+1\) independent families all meeting at a common boundary at \(x=1\) which \emph{can} support a single zero. This topology is illustrated in \cref{fig:block_topology}.

Let us also briefly comment on this observation in the context of other numerical bootstrap works. Unlike the strictly polynomial modular blocks, conformal blocks used in the conformal correlator bootstrap are exponentially damped for large scaling dimension. This asymptotic behavior results in functionals of finite order showing structure only up to some finite scaling dimension; see \cite{Chang:2025mwt} for more detailed discussion. The end result is that the large or infinite scaling dimension limits may be less important in those settings, but I should clarify that this is mostly speculation. In any case, the lack of closed-form expression for those blocks makes computing these asymptotic limits challenging.

\subsection{Modular blocks and numerics}

Now we turn to what these definitions of blocks mean for the numerics. The equation \eqref{eqn:truncated_crossing} with our scaled definiton of blocks is
\begin{equation}\label{eqn:rescaled_crossing}
  \vec{\widetilde{F}}_\mathbb{1}^N + \sum_{x,s} \widetilde{\rho}^N_{x,s} \vec{\widetilde{F}}_{x, s}^N = 0.
\end{equation}
Here we note that the rescaled blocks are a function of the number of functional components \(N\) which in turn is related to the highest derivative order of our truncation \(\Lambda\). For upgrading purposes, we will look at incrementing \(N\): assuming functional components are sorted by derivative order, for a given \(N\) the maximal derivative order is
\begin{equation}
  \Lambda(N) = 2\left\lceil\frac{\sqrt{8N+1}-1}{2}\right\rceil-1.
\end{equation}
As we add functional components, we will need to rescale our data, as the rescaled blocks themselves is depend on \(\Lambda(N)\).

We noted in the previous section that at \(x=1\) and/or \(s=\infty\) the blocks only are nonzero in the components with the maximal derivative order. This significantly restricts the space of independent vectors and in turn how many zeroes can be present. We already noted that there cannot be a fixed zero at \(x=1\) unless there is only one maximal-order vector component, in which case that is all there can be. More generally, we can only have as many zeroes as we have independent vectors, so the number of zeroes ``at infinity'' \(K_\text{fixed}^\infty+2K_\text{bulk}^\infty\) must be less than or equal to the number of components at the maximal derivative order.

As we discussed in prior sections, one of the ways in which positivity can be violated is if a fixed (single) zero turns into a double zero: from the perspective of the flow, an operator that was at the boundary of the a family of blocks becomes ``unfixed'' and moves into the bulk of that family. We have just seen that the 0th derivative for infinite twist and finite spin is actually identical for all finite spins. Each spin represents a family of blocks. If there is a fixed zero at \(x=1\) for one family, there is a fixed zero for all families. In other words, perhaps surprisingly, all finite-spin families of conformal blocks meet at \(x=1\) at a junction.

But now what if this fixed zero transmuted? If the first derivative were identical at \(x=1\) for all spin, then we would have quite a thorny problem: they would all develop double zeroes at the same moment. Which block family among \(s\in\{0,1\dots s_\text{max}\}\) would our unstuck zero move into? The fact that the 1st derivative \emph{does} depend on \(s\) means they will not all develop zeroes at once, and the fact that it's linear means only the families \(s=0,s_\text{max}\) need to be checked for transmutation.

\subsection{Upgrading from \(N=7\) to \(N=22\)}
\label{sec:results}

We can now finally turn to a specific bootstap problem that we upgrade. Start with the gap extremization problem in the \(c=5\) spinning modular bootstrap at \(N=7\) and upgrade to \(N=22\).\footnote{In the process of this work, I actually worked from \(N=5\) to \(N=35\). However, unusual small-\(N\) behavior before \(N=7\), uninteresting software problems above \(N=22\), and a general interest of clarity of presentation led me to choose to restrict to this range. I believe it to be representative of what upgrading flows to be nonetheless.} As we will see, even this apparently simple test case is technically quite challenging.

\begin{figure}[!htb]
  \centering
  \includegraphics[width=0.8\textwidth]{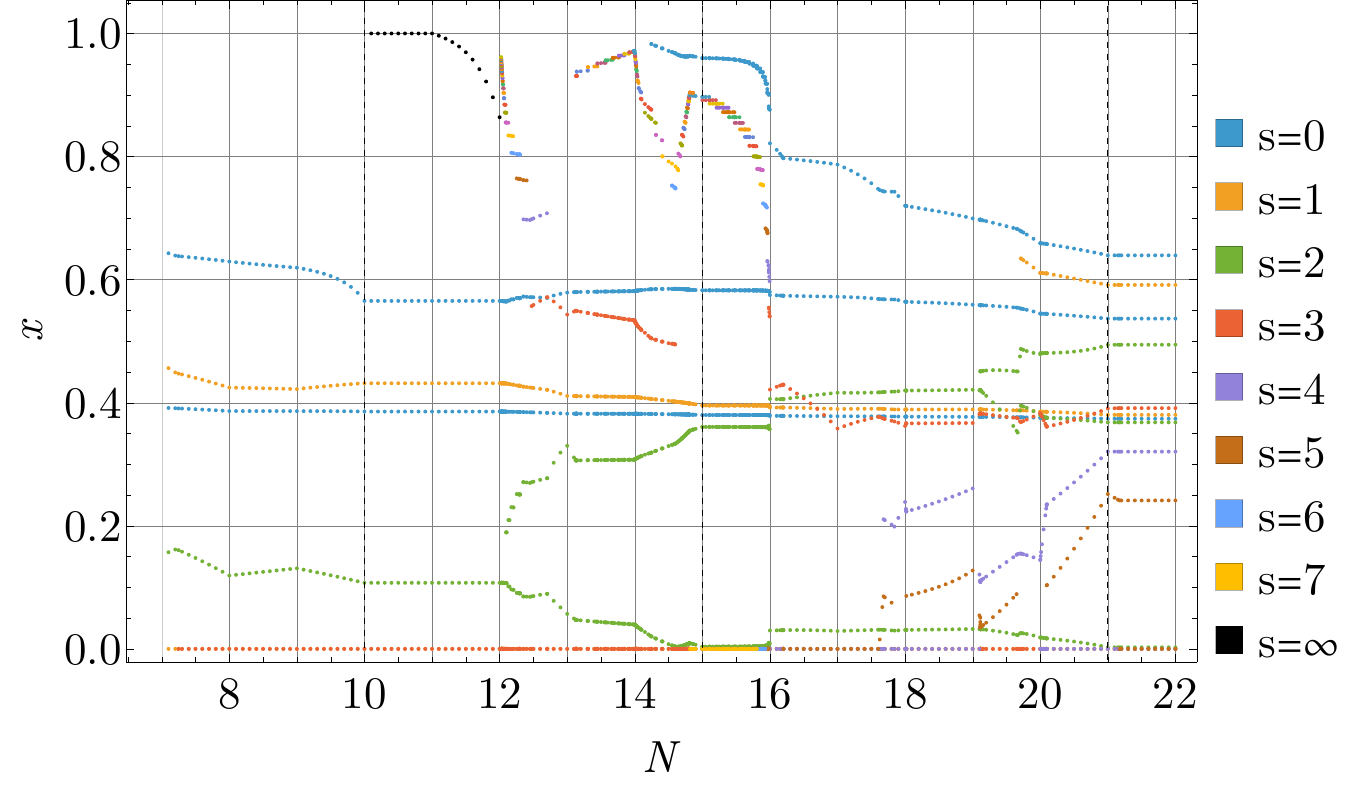}
  \caption{Plot of the compact twist \(x\), defined such that \(h=\frac{x}{1-x}\), of the various tracked operators in the spectrum as we flow from \(N=13\) to \(N=22\). Different colors correspond to different spins. The vertical dashed lines correspond to changes in maximum derivative order \(\Lambda\). The unlabeled multicolored feature between \(N=12\) and 16 are series of operators that appear and disappear for higher spins between \(s=3\) and \(s=30\) with all intermediate spins represented. See main text for further details.}
  \label{fig:flow-dimension}
\end{figure}
\begin{figure}[!htb]
  \centering
  \includegraphics[width=0.8\textwidth]{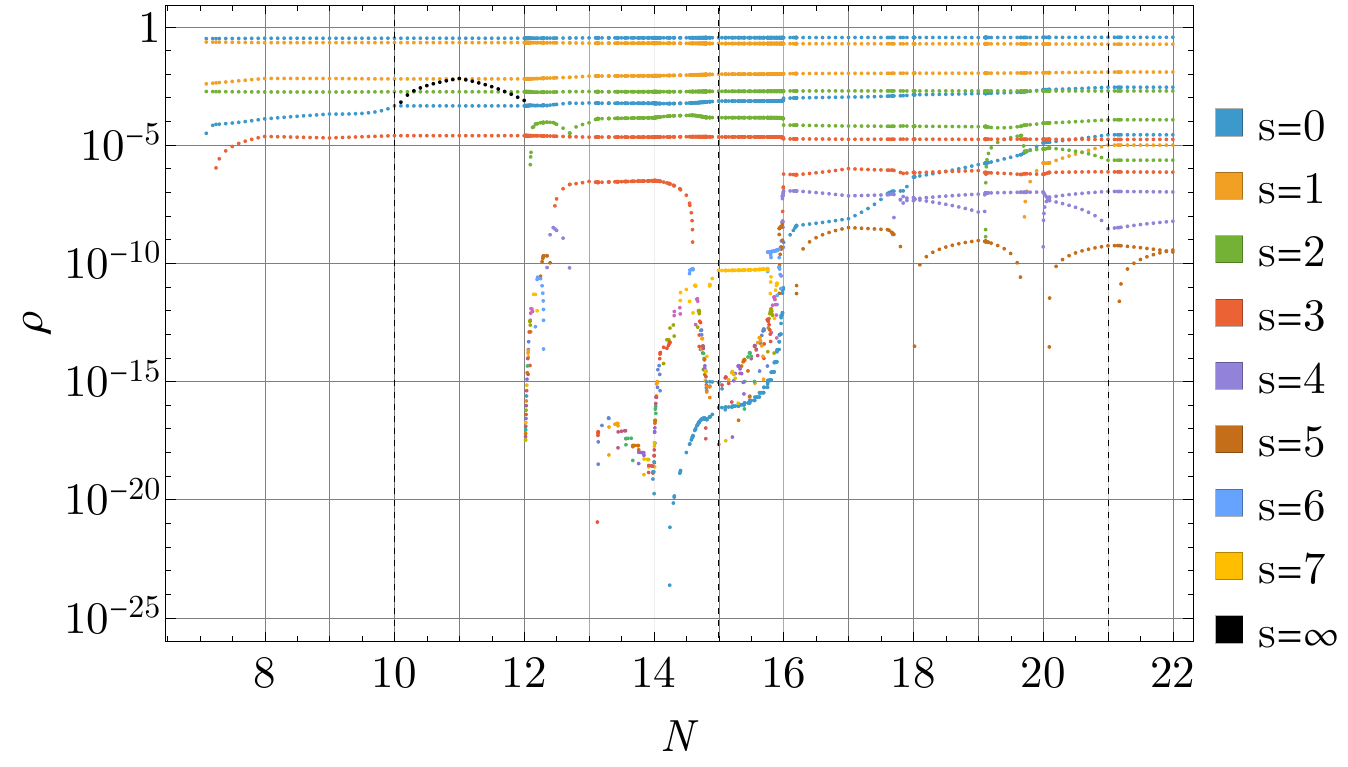}
  \caption{Plot of the coefficients \(\rho\) of the various tracked operators in the spectrum as we flow from \(N=13\) to \(N=19\). All coefficients are those that appear with the unscaled block components \(F_{h,\bar{h}}^{p,\bar p}\) except for \(s=\infty\). Different colors correspond to different spins. The vertical dashed lines correspond to changes in maximum derivative order \(\Lambda\). The multicolored set of points between \(N=12\) and 16 are series of operators that appear and disappear for higher spins between \(s=3\) and \(s=30\) with all intermediate spins represented. See main text for further details.}
  \label{fig:flow-coefficient}
\end{figure}

We present the summary of the upgrading procedure in two plots tracking our bulk operators' (compact) dimension and coefficient, respectively, along the flow to higher and higher \(N\) in \cref{fig:flow-dimension,fig:flow-coefficient}. We start at \(N=7\) with one bulk zero each in \(s=0,1,2\); however, we immediately hit upon a singular jacobian that deserves special attention, so we will save discussion for \cref{sec:step_jacobian}. After the initial roadblock, we proceed with very simple evolution in the dimension and spin of the operators until we get to \(N=10\). At \(N=10\), we hit upon another singular jacobian, and we have to add our first operator. As chance would have it it is an operator at infinity \(x=1\); as we go to \(N=11,12\) we get further operators at infinity, culminating in a cascade of operators on the way to \(N=13\) which we will discuss in \cref{sec:spin-hopping}.

Things proceed chaotically until \(N=16\), at which point the rest proceeds uneventfully all the way to \(N=22\) save for a few new operators for \(s=4,5\) and then \(s=1\).

Before we move on to discussing the excitement in more detail, let us remark upon some general observations. First of all, the coefficients appear fairly stable throughout the flow despite changing scaling dimensions. There are a few notable exceptions: the new \(s=0\) operator that appears at very high dimensions soon after \(N=14\) steadily grows as it descends to more reasonable dimensions, taking a long time to begin to stabilize.

Secondly, the flow is characterized by periods which mix the hectic and the humdrum. It's not obvious why any given new functional component would cause more or less drama; it would be worthwhile to see if there's some logic. Hints in this direction include the \(s=3\) bulk operator that appears between \(N=12\) and \(N=15\) before disappearing until shortly after \(N=16\), as well as the general trajectories of the \(s=4,5\) operators seemingly coinciding between \(N=18,19\) and \(N=20,21\).

\subsubsection{Stepping in \(\Lambda\), Jacobian singularity, and trivial functional components}
\label{sec:step_jacobian}
\begin{figure}
  \centering
  \includegraphics[width=0.8\textwidth]{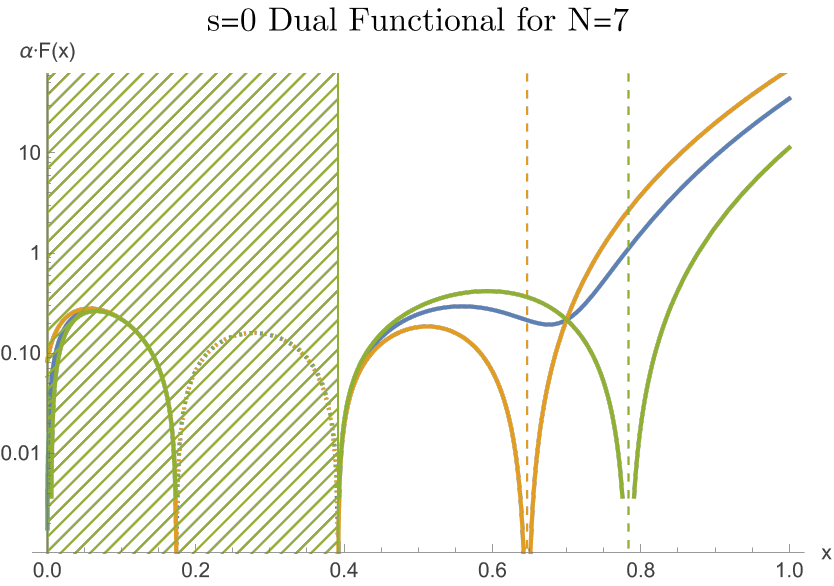}
  \caption{A plot overlaying the original \(N=7\) solution with a singular Jacobian and the two possible solutions that resolve that singularity. The dashed region to the left indicates the area below the gap where we don't care about the functional; the dotted curve there indicates a negative functional value (necessary, as this plot is a log plot). The blue solution is the original one, the orange solution is the correct one to continue a positive flow, and the green solution is another solution which immediately has a negative flow. The green solution provides the solution to flow into the direction of negative \(\beta\).}
  \label{fig:n7-jacobian}
\end{figure}

As noted, already in our very first step, from \(N=7\) to \(N=8\), we hit upon a challenge which we will now focus on. Following the discussion of section \ref{sec:jacobian_invertibility}, we see that our \(N=7\) solution has \(K_b=2\), \(K_f=1\), so if we were to upgrade,
\begin{equation}
  2K_b + K_f + 1 = 6
\end{equation}
which means that our \(N\) is immediately in violation of the bound in \cref{eqn:jacobian_n_bound}, meaning our Jacobian is immediately singular. Following the procedure to cure such a singularity, there are a priori two viable solutions for adding an operator. Only one of them does so while maintaining positivity as we flow from \(N=7\) to \(N=8\). The solutions are shown in \cref{fig:n7-jacobian}.

A similar situation occurs at \(N=10\); however, in that case, the solution is to add a fixed operator at the \(x=1\) junction (i.e. at \(\Delta=\infty\)), which is possible for \(N=11\) following the discussion in \cref{sec:block_topology}. However, this problem then repeats itself again upgrading to \(N=12\), where we must transmute our fixed operator at \(x=1\) to a bulk operator for \(s=\infty\). It repeats itself \emph{again} upgrading to \(N=13\) where at last the Jacobian is cured by introducing a new zero at finite spin and eventually eliminating the \(s=\infty\) operator entirely. Note that along the way, there is a long chain of operators that get added and removed in a sequence from very high spin.
\subsubsection{Spin-hopping is the performance killer}
\label{sec:spin-hopping}
Some of the most dramatic sequences happen with a specific chain of events: a sequence of operators that appear and disappear going from large to small spin before finally settling down.

\begin{figure}
  \centering
  \includegraphics[width=0.8\textwidth]{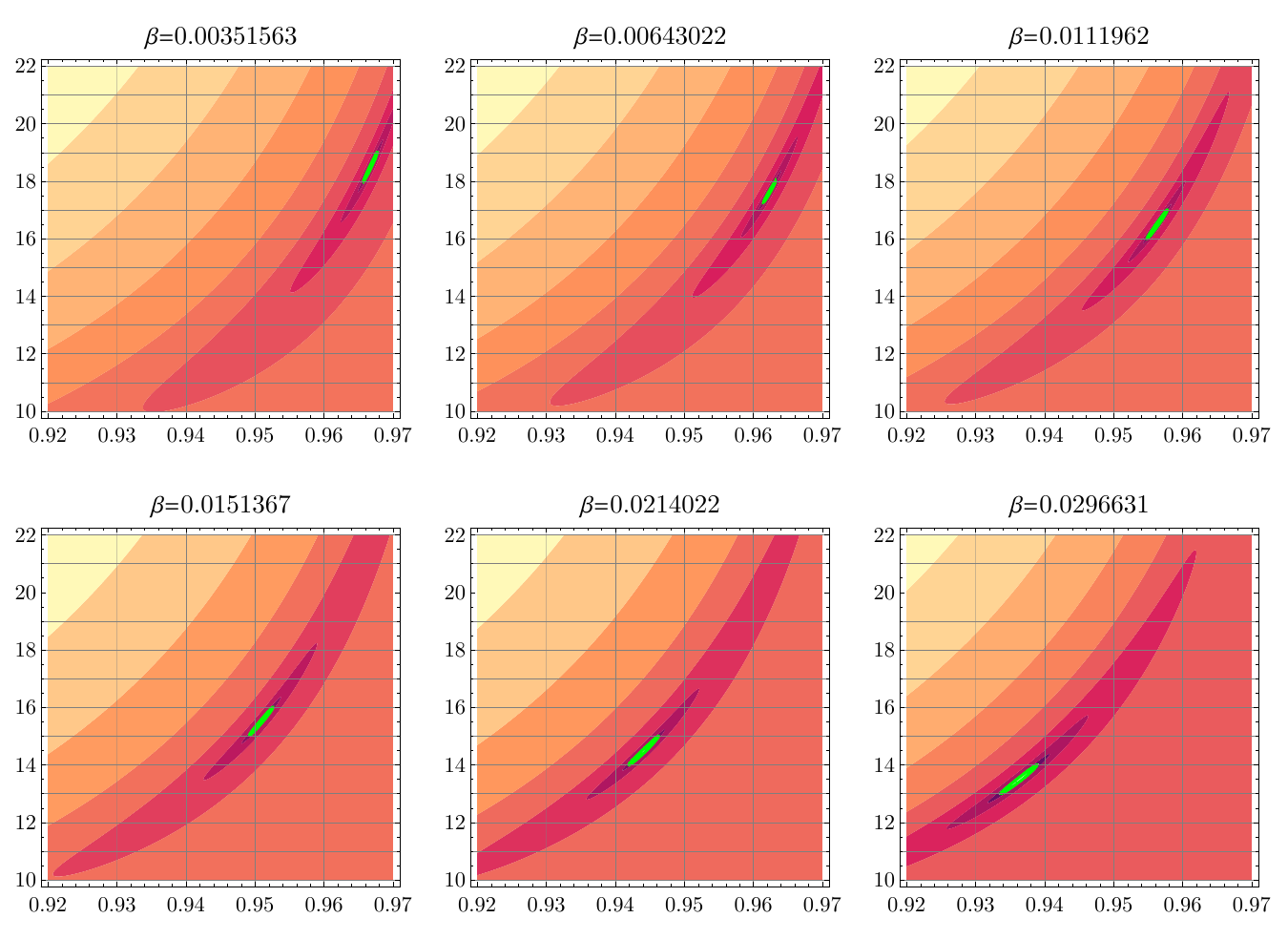}
  \caption{Plot of the functional \(\alpha\cdot F\) as a function of the compact dimension \(x\) and the spin \(s\) for various \(\beta\) in a flow from \(N=14\) to 15. The green contour is where the functional is zero; inside the closed curve the functional becomes negative. The contour touches at most two lines of integer spin.} \label{fig:spin-cascade}
\end{figure}

As we can see by eye, the operators that appear and disappear are clearly related to each other. We can get a sense of what's actually happening at the level of the functional by plotting the functional \(\alpha\cdot F\) for all \((\tau,s)\) in \cref{fig:spin-cascade}. We see that our functional becomes negative in the regions between integer spins, and that region of negativity steadily migrates from the higher spin regions to the lower spin regions, having to stop every time it must cross an integer spin as we need to ensure positivity. This is a tremendous performance killer, as for each of these well over a dozen crossings, we must stop and carefully solve for the branching point before modifying our equations to keep on the right path. This is made even worse if we have to cure Jacobian singularities at each of these points, as we have to in the cascade from \(N=12\) to \(N=13\). We can see the sheer density of points in these cascade regions; each of these corresponds to a point at which our flow had to stop and do some careful work. Compare this with the sparse and clean flows between \(N=8\) and \(N=10\), or even from \(N=16\) upward.

Why is this a problem? Ultimately, we would like an approach that is more efficient and effective in some situations than the general workhorse method based on convex optimization and \texttt{sdpb}. In both approaches, the algorithmic bottleneck will be inverting matrices to solve for our functional components. As we have the same number of functional components we are solving for, as a rule of thumb we can see that the scale of the matrices we must invert (in our case, our Jacobians) are roughly the same as those inverted in \texttt{sdpb}. Additionally, we must work with the same or similar high precision arithmetic, as the condition number will reflect the large scale between the smallest and largest coefficients, so we shouldn't expect any improvement there. The only way we can see an improvement is if we can reduce the number of these inversions. The number of iterations \texttt{sdpb} needs can vary, but for the cases of interest it tends to be between 100 and 1000 iterations. If we have to add and remove operators ~20-40 times and solving for each branching point takes ~10 root-finding steps, we see that we are already on the order of \texttt{sdpb}'s computations and see no significant algorithmic speed-up. Even if it isn't as bad as that, the prototype algorithm involves attempting to solve at a given step length in \(\beta\) and then backtracking which only makes this calculus worse.

\subsubsection{Speculations, directions not pursued}

As we have seen, the performance scaling of this approach does not seem to represent a significant algorithmic improvement over the mature and optimized \texttt{sdpb} approach for unitary CFTs. We might speculate that part of the problem is the choice of parametrization: we constrain our functional only with its zeroes at points for integer spins. But clearly there's more going on, with zeroes stretching along 1-dimensional families across dimension and spin as seen in \cref{fig:spin-cascade}. Could it be useful to use the language of algebraic varieties? Of course, we don't want to enforce positivity for non-integer spin, so it would seem to be quite a subtle line of inquiry.\footnote{I thank Nima Afkhami-Jeddi for discussions around this topic.} Another, more conservative solution to dealing with these spin cascades is to enforce nonnegativity for integer spins up to \(s_\text{max}\) as we have and then enforce nonnegativity for continuous spin from \(s_\text{max}\) to \(s=\infty\), which amounts to a relaxation of the optimization problem. We could then lower \(s_\text{max}\) significantly. In this context the only zeroes that can appear in the continuous spin region are 0-dimensional, but they can evolve in both spin and dimension. Perhaps this could be a fruitful direction for improvements.

One troubling feature is that it is not obvious a priori which functional component (i.e., which derivative) will have which effect. By eye, comparing functionals that are several components apart, we can see that they are not so dissimilar. This is in stark contrast to increments of 1 component which can range from being literally identical (with the extra component simply being 0) and being shockingly different (with the presence or absence of heavy and large-spin operators). We don't want components that are trivial as they will not improve our bounds, but if functionals vary so much in the short term (only for those variations to disappear in the final results) we will have wasted a lot of effort on the long and arduous path. I can only speculate that perhaps different orderings of the introduction of functional components will make for a clearer path, though it is not obvious to me that such an improved ordering exists.

Perhaps a way forward would be to use a more efficient and powerful functional basis than the computationally simple derivative basis. This could be some incorporation of the lessons of analytic functionals \cite{Mazac:2016qev,Mazac:2018ycv,Paulos:2019fkw,Paulos:2019gtx,Ghosh:2023onl}; I will confess a lack of expertise on the subject and the specific challenges that may arise.

Looking back to more traditional bootstrap methods, this work presents an interesting possibility. As our upgrading equations have a correspondence to a continuous family \(SDP_\beta\) of semidefinite programs, perhaps studying precisely these SDPs as functions of \(\beta\) using our mature and optimized \texttt{sdpb} solver could be interesting. In particular, there's no need to manually add and remove zeroes, removing the chief algorithmic problem, and it would allow for more efficient and effective hotstarting. Or perhaps this procedure could be used to better explore the space of solutions such as tracing the boundary of a bootstrap island more efficiently.

\section*{Acknowledgements}
This work originated as a collaboration with Aleix Gimenez-Grau and Slava Rychkov who contributed substantially throughout the early stages of prototyping and high-level algorithm design. I'm grateful for their effort, insight, and encouragement. I also thank Slava Rychkov specifically for further discussion and support through the writing process. I thank Nima Afkhami-Jeddi and Dalimil Maz\'a\v c for useful and insightful discussion on extremal flows and modular bootstrap. RSE is supported by Simons Foundation grant 915279 (IHES).

\appendix

\section{Prototype code availability}

The \textit{Mathematica} code used to compute the upgrading results in \cref{sec:results} is available at \url{https://gitlab.com/rajeev.erramilli/upgrading-extremal-flow-prototype/}; the latest commit at time of writing is \texttt{0d7c1c10a46170680c400a09af4b576a08bb5f1a}. Also included are the upgrading data used to make all of the plots. There are also the results from \texttt{sdpb} computations, but please note that they are only present to provide a starting point at \(N=7\) and to verify that the extremal flow is upgrading as we expect.

\section{More details on primal-dual solutions to conformal bootstrap problems}
\label{app:primal-dual}

Here I will elaborate on the precise technical details of primal-dual solutions to bootstrap problems that were elided over in \cref{sec:review}. Namely, \cref{eqn:identity_constraint} will more generally be expressed as follows for some set of conformal weights \(S\equiv \{(\Delta,s)_k\}\):

\begin{equation}\label{eqn:navigator_constraint}
  \vec\alpha\cdot\vec{F}_\mathbb{1} + \sum_{(\Delta,s)_k\in S} \rho_k\,\vec\alpha\cdot \vec{F}(\Delta_k,s_k)  = \vec\alpha \cdot \vec T = 0
\end{equation}

The term contracted with \(\vec\alpha\) is referred to as the ``target'' \(\mathbf{T}\) in \cite{El-Showk:2016mxr}, as it is the ``target'' of the remaining sum in \cref{eqn:primal}. Trivially when \(S=\emptyset\) this reduces to \cref{eqn:identity_constraint}, the expression for gap maximization problems. When \(S\) contains only a single weight \(\Delta_k\), this corresponds to the problem of extremizing the associated coefficient \(\rho_k\) (such as OPE coefficient optimization) up to subtleties of relative signs. More generally, the set \(S\) defines the vector \(\vec M\) and up to further subtleties to do with what exactly appears for \((\Delta_i,s_i)\) in \cref{eqn:primal-dual}, this is how the Navigator Function is formulated; see \cite{Reehorst:2021ykw} for the full and careful statement.

\small
\bibliographystyle{utphys}
\bibliography{mybib.bib}
\end{document}